\documentclass[a4paper,11pt]{article}

\usepackage[utf8]{inputenc}
\usepackage[numbers,sort&compress]{natbib}
\usepackage[margin=1.0in]{geometry}

\usepackage{graphicx}
\usepackage{amsmath}
\usepackage{hyperref}
\usepackage{placeins}
\usepackage{slashed}
\usepackage[colorinlistoftodos, shadow]{todonotes}
\usepackage{titlesec}
\usepackage{gensymb}
\usepackage{textcomp}
\usepackage{amssymb}
\usepackage{comment}

\def\refeq#1{\mbox{(\ref{#1})}}
\def\reffi#1{\mbox{Fig.~\ref{#1}}}
\def\reffis#1{\mbox{Figs.~\ref{#1}}}
\def\refta#1{\mbox{Tab.~\ref{#1}}}

\def\refse#1{\mbox{Sec.~\ref{#1}}}

\def\citere#1{\mbox{Ref.~\cite{#1}}}
\def\citeres#1{\mbox{Refs.~\cite{#1}}}

\def\ga{\gamma}

\newcommand{\GeV}{\unskip\,\mathrm{GeV}}
\newcommand{\MeV}{\unskip\,\mathrm{MeV}}
\newcommand{\TeV}{\unskip\,\mathrm{TeV}}

\def\mathswitch#1{\relax\ifmmode#1\else$#1$\fi}
\def\mathswitchr#1{\relax\ifmmode{\mathrm{#1}}\else$\mathrm{#1}$\fi}
\def\mathswitchit#1{\relax\ifmmode{#1}\else$#1$\fi}

\newcommand{\Pu}{\mathswitchr u}
\newcommand{\Pd}{\mathswitchr d}
\newcommand{\Ps}{\mathswitchr s}
\newcommand{\Pc}{\mathswitchr c}
\newcommand{\Pt}{\mathswitchr t}
\newcommand{\Pp}{\mathswitchr p}

\newcommand{\Pq}{\mathswitchit q}
\newcommand{\Pe}{\mathswitchr e}

\newcommand{\Pmu}{\mathswitchr {\mu}}

\newcommand{\PW}{\mathswitchr W}
\newcommand{\PZ}{\mathswitchr Z}

\newcommand{\PH}{\mathswitchr H}
\newcommand{\MZ}{\mathswitch {M_\PZ}}

\def\ie{i.e.\ }
\def\eg{e.g.\ }
\def\cf{cf.\ }

\newcommand{\ri}{{\mathrm{i}}}

\newcommand{\rd}{{\mathrm{d}}}

\newcommand{\fact}{{\mathrm{fact}}}

\newcommand{\U}{{\mathrm{U}}}

\def\sgn{\mathop{\mathrm{sgn}}\nolimits}

\newcommand{\Pqbar}{{\bar{\Pq}}}

\newcommand{\recola}{{\sc Recola}}
\newcommand{\collier}{{\sc Collier}}

\newcommand{\Pb}{\mathswitchr b}

\newcommand{\ppmmee}{\Pp \Pp \to \mu^+\mu^-\Pe^+\Pe^- +X}
\newcommand{\mmee}{\mu^+\mu^-\Pe^+\Pe^-}
\newcommand{\mmmm}{\mu^+\mu^-\mu^+\mu^-}
\newcommand{\mm}{\mu^+\mu^-}

\newcommand{\ppmmmm}{\Pp \Pp \to \mu^+\mu^-\mu^+\mu^- +X}

\newcommand{\mr}{\mathrm}

\def\bfi{\begin{figure}}
\def\efi{\end{figure}}

\def\draftdate{\relax}
\def\mda{\relax}
\def\mua{\relax}
\def\mla{\relax}
\def\draft{
\def\thtystars{******************************}
\def\sixtystars{\thtystars\thtystars}
\typeout{}
\typeout{\sixtystars**}
\typeout{* Draft mode!
         For final version remove \protect\draft\space in source file *}
\typeout{\sixtystars**}
\typeout{}
\def\draftdate{\today}
\def\mua{\marginpar[\boldmath\hfil$\uparrow$]%
                   {\boldmath$\uparrow$\hfil}%
                    \typeout{marginpar: $\uparrow$}\ignorespaces}
\def\mda{\marginpar[\boldmath\hfil$\downarrow$]%
                   {\boldmath$\downarrow$\hfil}%
                    \typeout{marginpar: $\downarrow$}\ignorespaces}
\def\mla{\marginpar[\boldmath\hfil$\rightarrow$]%
                   {\boldmath$\leftarrow $\hfil}%
                    \typeout{marginpar: $\leftrightarrow$}\ignorespaces}
\def\Mua{\marginpar[\boldmath\hfil$\Uparrow$]%
                   {\boldmath$\Uparrow$\hfil}%
                    \typeout{marginpar: $\uparrow$}\ignorespaces}
\def\Mda{\marginpar[\boldmath\hfil$\Downarrow$]%
                   {\boldmath$\Downarrow$\hfil}%
                    \typeout{marginpar: $\downarrow$}\ignorespaces}
\def\Mla{\marginpar[\boldmath\hfil$\Rightarrow$]%
                   {\boldmath$\Leftarrow $\hfil}%
                    \typeout{marginpar: $\leftrightarrow$}\ignorespaces}
\def\muua{\marginpar[\boldmath\hfil$\upuparrows$]%
                   {\boldmath$\upuparrows$\hfil}%
                    \typeout{marginpar: $\upuparrows$}\ignorespaces}
\def\mdda{\marginpar[\boldmath\hfil$\downdownarrows$]%
                   {\boldmath$\downdownarrows$\hfil}%
                    \typeout{marginpar: $\downdownarrows$}\ignorespaces}
\def\mlla{\marginpar[\boldmath\hfil$\leftleftarrows$]%
                   {\boldmath$\leftleftarrows $\hfil}%
                    \typeout{marginpar: $\leftleftarrows$}\ignorespaces}                    
\overfullrule 5pt
\oddsidemargin -15mm
\marginparwidth 29mm
}

\numberwithin{equation}{section}

\begin{document}

% --------------------- First Page ----------------------------------
\thispagestyle{empty}
\def\thefootnote{\fnsymbol{footnote}}
\setcounter{footnote}{1}
\null
\draftdate\hfill FR-PHENO-2016-013, ICCUB-16-036
\vfill
\begin{center}
  {\Large {\boldmath\bf {Next-to-leading-order electroweak corrections
        to the production of four charged leptons at the LHC}
\par} \vskip 2.5em
{\large
{\sc Benedikt Biedermann$^{1}$, Ansgar Denner$^{1}$, Stefan Dittmaier$^{2}$,\\ 
     Lars Hofer$^{3}$, Barbara J\"ager$^{4}$
}\\[2ex]
{\normalsize \it 
$^1$Julius-Maximilians-Universit\"at W\"urzburg, 
Institut f\"ur Theoretische Physik und Astrophysik, \\
97074 W\"urzburg, Germany
}\\[2ex]
{\normalsize \it 
$^2$Albert-Ludwigs-Universit\"at Freiburg, Physikalisches Institut, \\
79104 Freiburg, Germany
}\\[2ex]
{\normalsize \it 
$^3$Universitat de Barcelona (UB),
Departament de F\'{\i}sica Qu\`antica i Astrof\'{\i}sica (FQA),\\
Institut de Ci\`encies del Cosmo (ICCUB),
08006 Barcelona, Spain}
}\\[2ex]
{\normalsize \it
$^4$Eberhard Karls Universit\"at T\"ubingen, Institut f\"ur
Theoretische Physik, \\ 72076 T\"ubingen, Germany}
}
\par \vskip 1em
\end{center}\par
\vskip .0cm \vfill {\bf Abstract:} 
\par 
We present a state-of-the-art calculation of the next-to-leading-order
electroweak corrections to $\PZ\PZ$ production, including the leptonic
decays of the Z~bosons into $\mmee$ or $\mmmm$ final states.  We use
complete leading-order and next-to-leading-order matrix elements for
four-lepton production, including contributions of virtual photons and
all off-shell effects of Z~bosons, where the finite Z-boson width is
taken into account using the complex-mass scheme.  The matrix elements
are implemented into Monte Carlo programs allowing for the evaluation
of arbitrary differential distributions. We present integrated and
differential cross sections for the LHC at $13\TeV$ both for an
inclusive setup where only lepton identification cuts are applied, and
for a setup motivated by Higgs-boson analyses in the four-lepton decay
channel.  The electroweak corrections are divided into photonic and
purely weak contributions. The former show the well-known pronounced
tails near kinematical thresholds and resonances; the latter are
generically at the level of $\sim-5\%$ and reach several $-10\%$ in
the high-energy tails of distributions.  Comparing the results for
$\mmee$ and $\mmmm$ final states, we find significant differences
mainly in distributions that are sensitive to the $\mu^+\mu^-$ pairing
in the $\mmmm$ final state.  Differences between $\mmee$ and $\mmmm$
channels due to interferences of equal-flavour leptons in the final
state can reach up to $10\%$ in off-shell-sensitive regions.
Contributions induced by incoming photons, \ie photon--photon and
quark--photon channels, are included, but turn out to be
phenomenologically unimportant.
\par
\vskip 1cm
%\noindent
November 2016
\par
\null
\setcounter{page}{0}
\clearpage
\def\thefootnote{\arabic{footnote}}
\setcounter{footnote}{0}

%\tableofcontents

% --------------------- Main Part -----------------------------------
\section{Introduction}
\label{se:intro}

The physics programme of the LHC at Run~I was particularly successful
in the investigation of electroweak (EW) interactions and culminated
in the discovery of a Higgs boson, but no evidence for physics beyond
the Standard Model (SM) was found.  While the community is looking
forward to a major discovery at Run~II, an important task is the
precise measurement of the properties of the Higgs boson and the other
particles of the SM.  Small deviations from the predictions of the SM
in the observed event rates or distributions might reveal signs of new
physics.

One class of processes particularly relevant for tests of the EW
sector of the SM is EW gauge-boson pair production. These reactions
allow to measure the triple gauge-boson couplings and to study the EW
gauge bosons in more detail. Moreover, they constitute a background to
Higgs-boson production with subsequent decay into weak gauge-boson
pairs and to searches for new physics such as heavy vector bosons. In
the Higgs-signal region below the WW and ZZ production thresholds,
off-shell effects of the W and Z~bosons are of particular importance.
In this paper we focus on the production of Z-boson pairs with
subsequent decays to four charged leptons, covering all off-shell
domains in phase space.  While this channel has the smallest cross
section among the vector-boson pair production processes, it is the
cleanest, as it leads to final states with four charged leptons that
can be well studied experimentally.

At Run~I both ATLAS and CMS measured the cross section of Z-boson pair
production~\cite{Aad:2014wra,Chatrchyan:2012sga,Chatrchyan:2013oev,Aaboud:2016urj}
using final states with four charged leptons or two charged leptons
and two neutrinos.  The results of these measurements are in agreement
with the predictions of the SM and permitted to derive improved limits
on triple gauge-boson couplings between neutral gauge
bosons~\cite{Aad:2012awa,CMS:2014xja,Khachatryan:2015pba}.  Run~II
allows to improve these measurements, and first analyses have already
been published \cite{Aad:2015zqe,Khachatryan:2016txa}.

Precise measurements call for precise predictions.  The
next-to-leading order (NLO) QCD corrections to Z-boson pair production
were calculated a long time ago for stable Z~bosons
\cite{Ohnemus:1990za,Mele:1990bq} and including leptonic decays in the
narrow-width approximation \cite{Ohnemus:1994ff}. Once the one-loop
helicity amplitudes were available \cite{Dixon:1998py}, complete
calculations including spin correlations and off-shell effects became
possible \cite{Campbell:1999ah,Dixon:1999di}.  Gluon-induced one-loop
contributions were evaluated for stable Z~bosons
\cite{Dicus:1987dj,Glover:1988rg}, including off-shell effects
\cite{Matsuura:1991pj,Zecher:1994kb}, and studied as a background to
Higgs-boson searches \cite{Binoth:2008pr}.  NLO QCD corrections were
matched to parton showers in various frameworks with
\cite{Nason:2006hfa} and without
\cite{Hamilton:2010mb,Hoche:2010pf,Melia:2011tj,Frederix:2011ss}
including leptonic decays.  In \citere{Cascioli:2013gfa}, a
comprehensive NLO-QCD-based prediction was presented for off-shell
weak diboson production as a background to Higgs production.
Recently, the next-to-next-to-leading order (NNLO) QCD corrections to
Z-pair production have been calculated for the total cross section
\cite{Cascioli:2014yka} and including leptonic decays
\cite{Grazzini:2015hta}.  Although formally being beyond NNLO in the
pp~cross section, even the NLO corrections to the loop-induced
gluon-fusion channel were
calculated~\cite{Caola:2015psa,Caola:2016trd,Alioli:2016xab} because
of their particular relevance in Higgs-boson analyses.

Besides QCD corrections also EW NLO corrections are important for
precise predictions of vector-boson pair production at the LHC. EW
corrections typically increase with energy owing to the presence of
large Sudakov and other subleading EW logarithms
\cite{Beenakker:1993tt,Beccaria:1998qe,Ciafaloni:1998xg,Kuhn:1999de,Fadin:1999bq,Denner:2000jv}
and reach several $10\%$ in the high-energy tails of distributions. In
addition, photonic corrections lead to pronounced radiative tails near
resonances or kinematical thresholds.  Logarithmic EW corrections to
gauge-boson pair production at the LHC were studied in
\citere{Accomando:2004de} and found to reach 30\% for Z-pair
production for ZZ invariant masses in the TeV range.  Later, the
complete NLO EW corrections were calculated for stable vector bosons
and all pair production processes including photon-induced
contributions \cite{Bierweiler:2013dja,Baglio:2013toa}.  The size and
in particular the non-uniform effect on the shapes of distributions
were confirmed. Leptonic vector-boson decays were first included in
NLO EW calculations in the form of a consistent expansion about the
resonances for W-pair production~\cite{Billoni:2013aba}, and in an
approximate variant via the {\tt Herwig++} Monte Carlo generator for
WW, WZ, and ZZ production~\cite{Gieseke:2014gka}.  Most recently, NLO
EW calculations based on full $2\to4$ particle amplitudes, including
all off-shell effects, have been presented for
W-pair~\cite{Biedermann:2016guo} and Z-pair
production~\cite{Biedermann:2016yvs} for four-lepton final states of
different fermion generations (\ie without identical particle effects
or WW/ZZ interferences).  For Z-pair production, the off-shell effects
include also the contributions of virtual photons that cannot be
separated from the $\PZ$-pair signal, but only suppressed by using
appropriate invariant-mass cuts.  Note that these full off-shell
calculations are essential to safely assess the EW corrections below
the WW and ZZ thresholds, \ie in the kinematical region where
WW$^*$/ZZ$^*$ production appears as background to Higgs-boson
analyses.  Moreover, a detailed comparison of the full four-lepton
calculation~\cite{Biedermann:2016guo} to the double-pole approximation
for W-boson pairs~\cite{Billoni:2013aba} revealed limitations of the
latter approach for transverse-momentum distributions of the leptons
in the high-energy domain where new-physics signals are searched for.

In \citere{Biedermann:2016yvs} we have presented some selected results
for the NLO EW corrections to off-shell ZZ production in a scenario
relevant for Higgs-boson studies.  In this paper we provide more
detailed phenomenological studies in various phase-space regions
relevant for LHC analyses for $\ppmmee$ and completely new results on
$\ppmmmm$, including interference effects from identical final-state
leptons. We follow the same concepts and strategies as in
\citeres{Biedermann:2016yvs,Biedermann:2016guo}, \ie finite-width
effects of the Z~bosons are consistently included using the
complex-mass scheme \cite{Denner:1999gp,Denner:2005fg,Denner:2006ic},
so that we obtain NLO EW precision everywhere in phase space.  We also
include photon-induced partonic processes originating from $\ga\ga$ or
$\Pq\ga/\Pqbar\ga$ initial states.

The paper is organized as follows: Some details on the calculational
methods are presented in \refse{se:methods}. Phenomenological results
for two different experimental setups are discussed in
\refse{se:results}. Our conclusions are given in \refse{se:conclusion}.

\section{Details of the calculation}\label{se:methods}

\subsection{Partonic channels} 

\begin{figure}
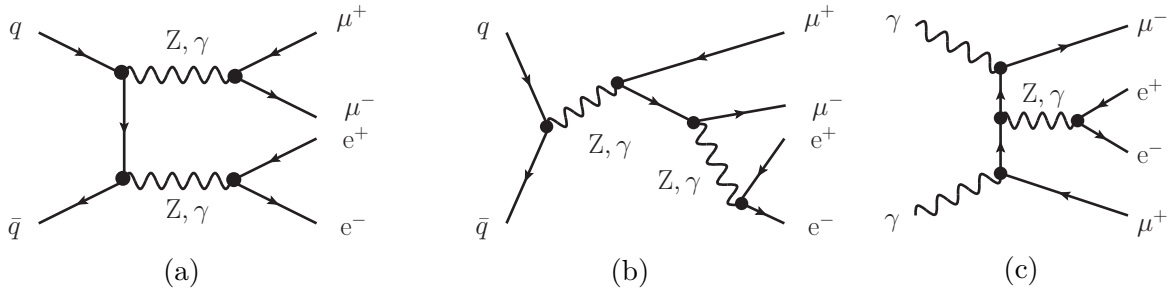

\begin{center}
    \begin{minipage}{0.33\textwidth}
      \includegraphics[width=\textwidth]{diagrams/{{born_QQ4lep}}}
      \begin{center} \vspace{-5mm} (a)~~~~~ \end{center}
    \end{minipage}
    \qquad
    \begin{minipage}{0.33\textwidth}
      \includegraphics[width=\textwidth]{diagrams/{{born_QQ4lepBackground}}}
      \begin{center} \vspace{-5mm} (b)~~~~~~~~~ \end{center}
    \end{minipage}
    \begin{minipage}{0.26\textwidth}
      \includegraphics[width=\textwidth]{diagrams/{{born_phph4lep}}}
      \begin{center} \vspace{-5mm} (c)~~~~ \end{center}
    \end{minipage}
\end{center}
\caption{Sample tree-level diagrams contributing at $O(\alpha^4)$. The dominant $\bar q q$ channel (a,b) defines the LO contribution, while the photon-induced $\gamma\gamma$ channel (c) is counted as a correction.}\label{fig:born}
\end{figure}

The leading-order (LO) cross sections of the two processes
$\Pp\Pp\to\mmee+X$ and $\Pp\Pp\to\mmmm+X$ receive contributions from
the quark--antiquark annihilation channels
\begin{align}
 \bar q  q/q \bar q & \to \mu^+\mu^-\Pe^+\Pe^-,\;\mu^+\mu^-\mu^+\mu^-, 
\label{eq:LOdefinition}
\end{align} 
with $q\in\{\Pu,\Pd,\Pc,\Ps,\Pb\}$.  Sample diagrams for these
channels, which are generically called $\bar qq$ channels in the
following, are shown in \reffis{fig:born}(a) and \ref{fig:born}(b).
Note that all LO diagrams involve $\PZ$-boson and photon exchange only.
There are LO channels 
with two photons in the initial state as well,
\begin{align}
\gamma\gamma &\to \mu^+\mu^-\Pe^+\Pe^-,\;\mu^+\mu^-\mu^+\mu^-, \label{eq:phophoBorn}
\end{align}
with corresponding diagrams shown in \reffi{fig:born}(c). 
Due to their small numerical impact, verified in \refse{se:results}, 
we consider their contribution as a
correction $\delta_{\gamma\gamma}$ to the $\bar qq$-induced LO cross
section and do not include higher-order corrections to these
processes.  

The NLO EW corrections comprise virtual and real contributions of the
$\bar qq$ channels,
\begin{align}
  \bar q  q/q \bar q &\to \mu^+\mu^-\Pe^+\Pe^- \,(+\gamma),\;\mu^+\mu^-\mu^+\mu^- \,(+\gamma),\label{eq:qqNLO}
\end{align}
and the real photon-induced contributions with one (anti)quark and one
photon in the initial state,
\begin{align}
  \gamma q/q \gamma &\to \mu^+\mu^-\Pe^+\Pe^- \,q,\;\mu^+\mu^-\mu^+\mu^- \,q, \nonumber\\
  \gamma \bar q/\bar q \gamma &\to \mu^+\mu^-\Pe^+\Pe^-\, \bar q,\;\mu^+\mu^-\mu^+\mu^-\, \bar q,\label{eq:qphoNLO}
\end{align} 
generically referred to as $q\gamma$ channels in the following.

\subsection{Virtual corrections}

\begin{figure}
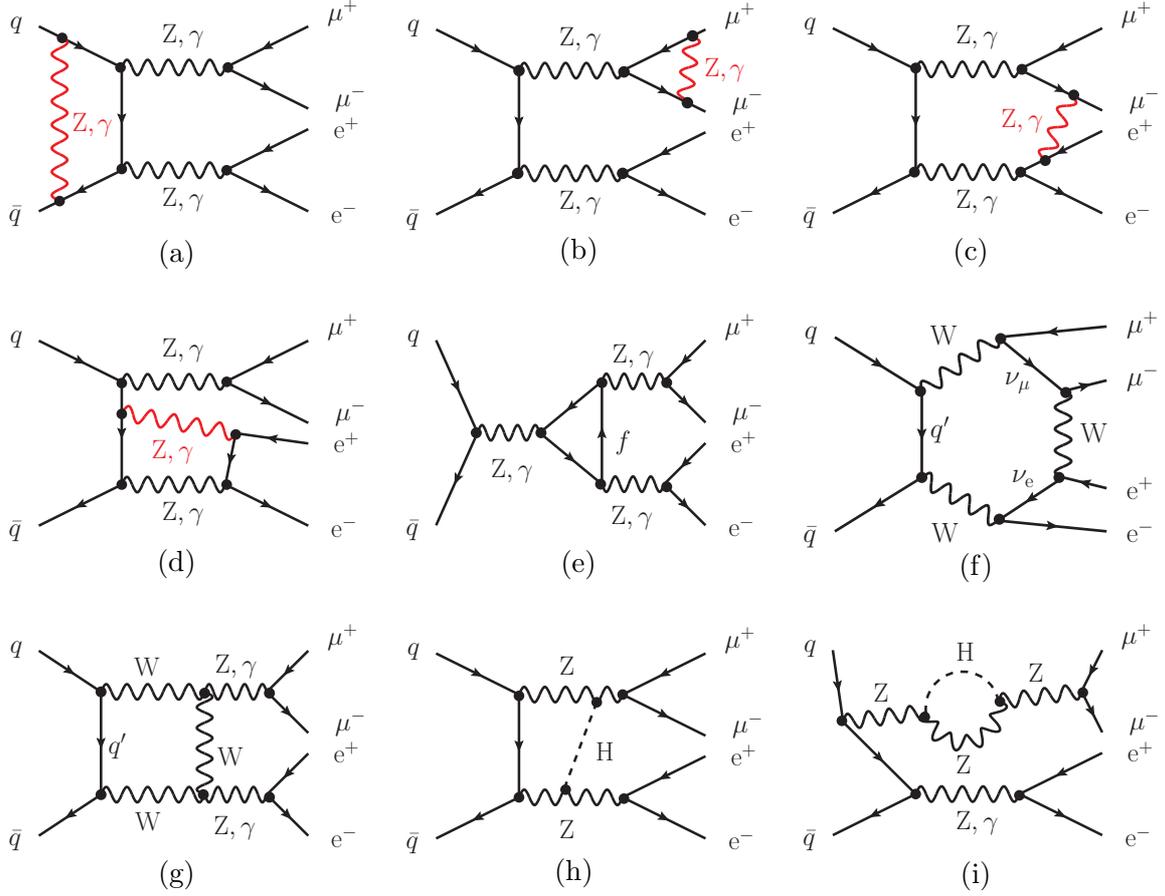

  \begin{center}
    \begin{minipage}{0.32\textwidth}
      \includegraphics[width=\textwidth]{diagrams/{{QQ4lepNLOEW1}}}
      \begin{center} \vspace{-5mm} (a)~~~~~ \end{center}
    \end{minipage}
%    \qquad
    \begin{minipage}{0.32\textwidth}
      \includegraphics[width=\textwidth]{diagrams/{{QQ4lepNLOEW3}}}
      \begin{center} \vspace{-5mm} (b)~~~~ \end{center}
    \end{minipage}
    \begin{minipage}{0.32\textwidth}
      \includegraphics[width=\textwidth]{diagrams/{{QQ4lepNLOEW4}}} 
      \begin{center} \vspace{-5mm} (c)~~~~~ \end{center}
    \end{minipage}
  \end{center} 
    \begin{center}
    \begin{minipage}{0.32\textwidth}
      \includegraphics[width=\textwidth]{diagrams/{{QQ4lepNLOEW2}}} 
      \begin{center} \vspace{-5mm} (d)~~~~~ \end{center}
    \end{minipage}
    \begin{minipage}{0.32\textwidth}
      \includegraphics[width=\textwidth]{diagrams/{{QQ4lepNLOEW9}}}
      \begin{center} \vspace{-5mm} (e)~~~~ \end{center}
    \end{minipage}
    \begin{minipage}{0.32\textwidth}
      \includegraphics[width=\textwidth]{diagrams/{{QQ4lepNLOEW5}}}
      \begin{center} \vspace{-5mm} (f)~~~~ \end{center}
    \end{minipage}
  \end{center}
  \begin{center}
    \begin{minipage}{0.32\textwidth}
      \includegraphics[width=\textwidth]{diagrams/{{QQ4lepNLOEW10}}} 
      \begin{center} \vspace{-5mm} (g)~~~~~ \end{center}
    \end{minipage}
    \begin{minipage}{0.32\textwidth}
      \includegraphics[width=\textwidth]{diagrams/{{QQ4lepNLOEW8}}} 
      \begin{center} \vspace{-5mm} (h)~~~~~ \end{center}
    \end{minipage}
    \begin{minipage}{0.32\textwidth}
      \includegraphics[width=\textwidth]{diagrams/{{QQ4lepNLOEW6}}}
      \begin{center} \vspace{-5mm} (i)~~~~ \end{center}
    \end{minipage}
  \end{center}
\caption{Sample diagrams for the virtual EW corrections. Diagram types
  (a)--(d) are obtained by photon and Z-boson insertions between the
  fermion lines of the tree-level diagrams in \reffi{fig:born}(a).
  The remaining diagrams may involve couplings (f)--(i) or corrections
  to vertices (e) that are not present at LO.}
\label{fig:virtdiagrams}
\end{figure}
\begin{figure}
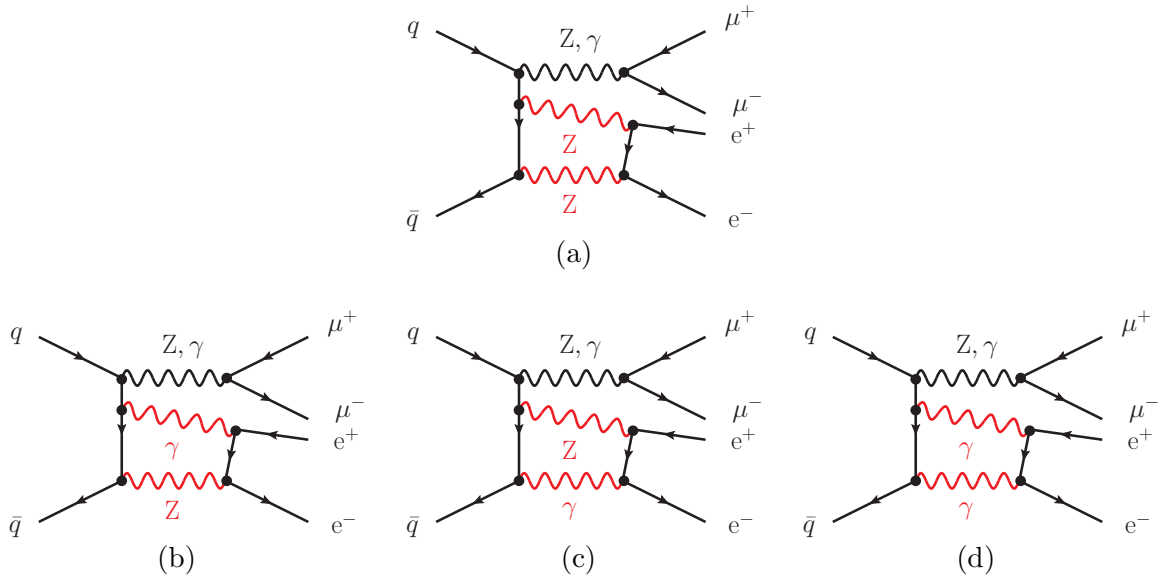

  \begin{center}
    \begin{minipage}{0.32\textwidth}

      \includegraphics[width=\textwidth]{diagrams/{{QQ4lepNLOEW2_weak}}}
      \begin{center} \vspace{-5mm} (a)~~~~~ \end{center}
    \end{minipage}
  \end{center} 
  \begin{center}
    \begin{minipage}{0.32\textwidth}
      \includegraphics[width=\textwidth]{diagrams/{{QQ4lepNLOEW2_qed1}}} 
      \begin{center} \vspace{-5mm} (b)~~~~~ \end{center}
    \end{minipage}
    \begin{minipage}{0.32\textwidth}
      \includegraphics[width=\textwidth]{diagrams/{{QQ4lepNLOEW2_qed2}}}
      \begin{center} \vspace{-5mm} (c)~~~~ \end{center}
    \end{minipage}
    \begin{minipage}{0.32\textwidth}
      \includegraphics[width=\textwidth]{diagrams/{{QQ4lepNLOEW2_qed3}}}
      \begin{center} \vspace{-5mm} (d)~~~~ \end{center}
    \end{minipage}  \end{center}
\caption{Illustration of the splitting of EW corrections into purely weak (a) and photonic (b)--(d)
contributions for the diagram type shown in \reffi{fig:virtdiagrams}(d).}\label{fig:weakPhotonicSplitting}
\end{figure}
The one-loop virtual corrections to the $\bar q q$ channels are
computed including the full set of Feynman diagrams.  We employ the
complex-mass scheme for the proper handling of unstable internal
particles~\cite{Denner:1999gp,Denner:2005fg,Denner:2006ic}.  This
approach allows the simultaneous treatment of phase-space regions
above, near, and below the Z~resonances within a single framework,
leading to NLO accuracy both in resonant and non-resonant regions.
Sample diagrams for the virtual EW corrections are shown in
\reffi{fig:virtdiagrams}.  A first set of diagrams is obtained by
exchanging Z~bosons or photons in all possible ways between the
fermion lines of the tree-level diagrams in \reffi{fig:born}: Diagram
types (a) and (b) of \reffi{fig:virtdiagrams} would also be present in
narrow-width or pole approximations for the Z~bosons and contain
separate corrections to the production and the decay of the Z~boson.
Diagrams (c) and (d) feature correlations between the initial and
final states or between different Z-boson decays and are only present
in a full off-shell calculation.  The sample diagrams (e)--(i) cannot
be obtained by naive vector-boson insertions between fermion lines.
They involve, for example, closed fermion loops (e) or the exchange of
W or Higgs bosons.

In our calculation, we perform a gauge-invariant decomposition of the
full NLO EW corrections $\Delta\sigma_{\rm NLO}$ into a purely weak
part $\Delta\sigma_{\rm NLO}^{\rm weak}$ and a photonic part
$\Delta\sigma_{\rm NLO}^{\rm phot}$.  The virtual photonic part is
defined as the set of all diagrams with at least one photon in the
loop coupling to the fermion lines.  The weak contribution is then the
set of all remaining one-loop diagrams, including also self-energy
insertions and vertex corrections induced by closed fermion loops.
The contributions to the renormalization constants have to be split
accordingly. This splitting is possible, because the LO process with
four charged leptons in the final state does not involve
charged-current interactions, \ie there is no W-boson exchange at tree
level.  The vector-boson insertions between fermion lines exemplarily
shown in the diagrams (a)--(d) of \reffi{fig:virtdiagrams} thus
exhaust all generic possibilities how a photon appears in a loop
propagator and can systematically be used to construct the virtual
photonic contribution.  Figure~\ref{fig:weakPhotonicSplitting} shows
the decomposition of the eight diagrams represented by
\reffi{fig:virtdiagrams}(d) into the purely weak part with only
Z~bosons in the loop coupling to fermion lines (upper row) and the
photonic part with one or two photons in the loop (lower row).  Note
that the criterion for the splitting considers only the vector bosons
in the loop, while it does not refer to the tree-level part of the
diagram.  The contributions to the field renormalization constants of
the fermions are decomposed in an analogous manner.  Since only loops
with internal photons lead to soft and collinear divergences, the
purely weak contribution is infrared (IR) finite. The full and finite
photonic corrections to the $\bar qq$ channels, on the other hand,
comprise the virtual photonic corrections plus the real photon
emission described in the next section.

In general, the decomposition of an amplitude into gauge-invariant
parts requires great caution.  The gauge-invariant isolation of
photonic corrections in processes that proceed in LO via
neutral-current interactions is only possible because the LO amplitude
can be interpreted as a result obtained in a $\U(1)_\gamma\times
\U(1)_Z$ gauge theory with the same fermion content and the same
couplings as the SM.  Here, $\U(1)_\gamma$ refers to the
electromagnetic gauge group with the photon as massless gauge boson
and $\U(1)_Z$ to an Abelian gauge theory with the Z~boson as gauge
boson.  The corrections within this theory, which is a consistent and
renormalizable field theory on its own (see \eg \citere{Piguet:1974tp}
or \citere{Collins:1984xc}, Chapter 12.9), form a gauge-invariant
subset of the full SM corrections to our neutral-current process.
Since the electric charges of the fermions are free, independent
parameters in the $\U(1)_\gamma\times \U(1)_Z$ theory, the photonic
contributions form a gauge-invariant subset of the corrections (which
could even be further decomposed into subsets defined by the global
charge factors of the fermions linked by the photon).  Note that the
subset of closed fermion loops could be isolated as another
gauge-invariant part of the EW correction, since each fermion
generation (in the absence of generation mixing) delivers a
gauge-invariant subset of diagrams. For the process at hand, we,
however, prefer to keep the closed fermion loops in the weak
corrections.

\subsection{Real corrections}\label{sec:realEmission}

\begin{figure}
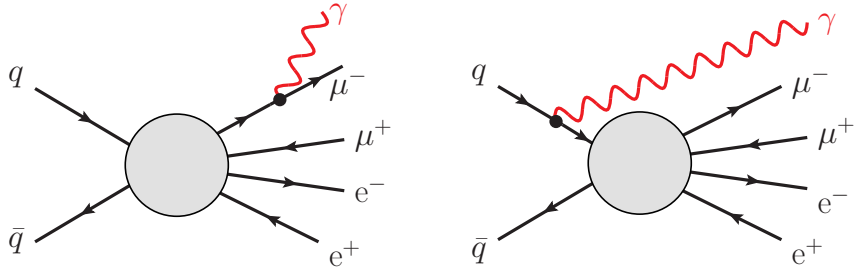

  \begin{center}
    \begin{minipage}{0.35\textwidth}
      \includegraphics[width=\textwidth]{diagrams/{{RealGammaFS}}} 
    \end{minipage}
    \quad
    \begin{minipage}{0.35\textwidth}
      \vspace{1mm}
      \includegraphics[width=\textwidth]{diagrams/{{RealGammaIS}}}
    \end{minipage}
  \end{center}
  \caption{Illustration of real photon radiation off the final-state leptons and off the initial-state quarks.
  }\label{fig:realGammaEmission}
\end{figure}

The real corrections to the $\bar qq$ channels include all possible
ways of photon radiation off the initial-state quarks or off the
final-state leptons, schematically depicted in
\reffi{fig:realGammaEmission}.  The phase-space integrations over the
squared real-emission amplitudes diverge if the radiated photon
becomes soft or collinear to one of the external fermions.  For
IR-safe (\ie soft- and collinear-safe) observables, however, the
collinear final-state singularities and the soft singularities cancel
exactly the corresponding soft and collinear divergences from the
virtual corrections after integration over the phase space.  The
collinear initial-state singularities do not fully cancel between real
and virtual corrections; the remnants are absorbed into the parton
distribution functions (PDFs) via factorization, in complete analogy
to the usual procedure applied in QCD.

We employ the dipole subtraction formalism for the numerical
integration of the real corrections.  In detail, we apply two
different variants based on dimensional regularization
\cite{Catani:1996vz} and mass regularization \cite{Dittmaier:1999mb},
respectively.  The results obtained with the two approaches are in
perfect agreement.  The underlying idea is to add and subtract
auxiliary terms $|{\cal{M}}_{\rm sub}|^2$ at the integrand level which
pointwise mimic the universal singularity structure of the squared
real matrix elements $|{\cal{M}}_{\rm real}|^2$ and which, on the
other hand, can be integrated analytically in a process-independent
way. In the dipole subtraction approach, the subtraction function is
constructed from ``emitter--spectator pairs'', where the ``emitter''
is the particle whose collinear splitting leads to an IR singularity
and the ``spectator'' is the particle balancing momentum and charge
conservation in the emission process.  Schematically, the NLO
correction $\Delta\sigma_{\rm NLO}$ then reads
\begin{align}
 \hspace{-1mm}\Delta\sigma_{\rm NLO} = &\int_{n+1} {\rm d}\Phi\Big[|{\cal{M}}_{\rm real}|^2-|{\cal{M}}_{\rm sub}|^2\Big]  
 + \int_{n}{\rm d}\tilde\Phi\left[\left(\int_1 [{\rm d}k]|{\cal{M}}_{\rm sub}|^2 \right)+ 
 2{\rm Re}\left[{\cal{M}}^*_{\rm Born}{\cal{M}}_{\rm virt}\right]\right]
\nonumber\\
& {} + \int_0^1\rd x \int_{n} {\rm d}\Phi(x)\,
P_{\fact}(x)\,|{\cal{M}}_{\rm Born}|^2,
\label{dNLO-masterFormula}
\end{align} 
where $2{\rm Re}\left[{\cal{M}}^*_{\rm Born}{\cal{M}}_{\rm
    virt}\right]$ denotes the interference of the Born amplitude
${\cal{M}}_{\rm Born}$ with the one-loop amplitude ${\cal{M}}_{\rm
  virt}$.  The last term represents the (IR-divergent) factorization
contribution from the PDF redefinition, which takes 
the form of a
convolution over the momentum fraction $x$ quantifying the momentum
loss via collinear parton/photon emission.  In order to achieve the
form of Eq.~\refeq{dNLO-masterFormula}, the phase-space measure ${\rm
  d}\Phi$ of the $(n+1)$-particle real phase space, which includes the
bremsstrahlung photon, is decomposed into a ``reduced $n$-particle
phase space'' ${\rm d}\tilde\Phi$ without photon emission and the
remaining one-particle phase space $[{\rm d}k]$ of the bremsstrahlung
photon according to ${\rm d}\Phi = {\rm d}\tilde\Phi\times[{\rm d}k]$.
The integration over $[{\rm d}k]$ involves an integration over the
variable $x$ which controls the momentum loss from initial-state
radiation.  Splitting off the soft singularities developing in this
integration of the subtraction terms over $x$, decomposes the integral
$\int_1 [{\rm d}k]|{\cal{M}}_{\rm sub}|^2$ into a part proportional to
$\delta(1-x)$ and a continuum part.  The former contribution can be
analytically combined with the virtual corrections, the latter with
the factorization contribution, to produce individually IR-finite
terms which may be integrated separately in a fully numerical way:
\begin{align}
  \Delta\sigma_{\rm NLO} =&\int_{n+1} {\rm d}\Phi\Big[|{\cal{M}}_{\rm real}|^2\Theta(p_1,\ldots,p_n,p_\gamma)-|{\cal{M}}_{\rm sub}|^2\Theta(\tilde p_1,\ldots,\tilde p_n)\Big] \nonumber\\
 & + \int_{n}{\rm d}\tilde\Phi\int_0^1{\rm d}x\left[|{{\cal{M}}_{\rm sub}(x)}|^2_{\rm fin} + \delta(1-x)2{\rm Re}\left[{{\cal{M}}^*_{\rm Born}{\cal{M}}_{\rm virt}}\right]_{\rm fin} \right]\Theta(\tilde p_1,\ldots,\tilde p_n).\label{dNLOqqfinite}
\end{align}
Here, $|{{\cal{M}}_{\rm sub}(x)}|^2_{\rm fin}$ and 
$2{\rm Re}\left[{{\cal{M}}^*_{\rm Born}{\cal{M}}_{\rm virt}}\right]_{\rm fin}$ 
are the finite parts resulting from this splitting of the $x$ integration
into continuum and endpoint parts.
The momenta $\tilde p_i$ of the reduced $n$-particle phase space in
the first integral are related to the momenta $p_i$ of the 
$(n+1)$-particle phase space such that $\tilde p_k\to p_k+p_\gamma$ for
the emitter $k$ in the collinear limit $p_k \cdot p_\gamma\to0$.
The integration over $x$ is a remainder from the factorized
one-particle phase space and is only present for radiation off
initial-state particles.  The $\Theta(p_1,\dots)$~functions represent
the phase-space cuts applied to the particle momenta $\{p_1,\dots\}$,
possibly after applying a
recombination procedure which is discussed in the next paragraphs.

\paragraph{Quark-induced channels---the collinear-safe setup:}

Observables that are collinear safe with respect to final-state
radiation---as described in \citere{Dittmaier:1999mb}---are
constructed by applying an appropriate procedure for recombining
radiated photons with nearly collinear final-state leptons.  In
collinear regions, a photon--lepton pair with photon and lepton
momenta $p_\gamma$ and $p_\ell$ is treated as one quasi-particle with
momentum $p_{\gamma\ell}=p_\gamma+p_\ell$ if the two-particle
separation $\Delta R_{\gamma,\ell}$, as defined in
Eq.~\refeq{eq:recombi} below, is beneath a given threshold.  Any
phase-space cut or any evaluation of an observable is performed for
the recombined momenta, while the matrix elements themselves are
evaluated with the original kinematics.  Both the local and the
integrated subtraction terms and the virtual corrections are cut in
terms of the $n$-particle kinematics.  Note that the difference
$|{\cal{M}}_{\rm
  real}|^2\Theta(p_1,\ldots,p_n,p_\gamma)-|{\cal{M}}_{\rm
  sub}|^2\Theta(\tilde p_1,\ldots,\tilde p_n)$ is only integrable
after photon recombination, because this procedure ensures that the
two $\Theta$ functions become equal in the soft/collinear regions (up
to edge-of-phase-space effects which do not spoil integrability).
Since collinear lepton--photon configurations are treated fully
inclusively within some collinear cone defined by the photon
recombination, the conditions for the KLN theorem are fulfilled,
guaranteeing the cancellation of the collinear mass singularity.  The
formation of such a quasi-particle is close to the experimental
concept of ``dressed leptons'', as \eg described by the ATLAS
collaboration in \citere{Aad:2011gj}.

\paragraph{Quark-induced channels---the collinear-unsafe setup:}

It is not a priori necessary that observables sensitive to photon
radiation off a final-state lepton are defined in a collinear-safe
way.  The reason is that photons and charged leptons may be detected
in geometrically separated places, \ie the photons in the
electromagnetic calorimeter and the muons in the muon chamber.  This
allows the measurement of an arbitrarily collinear photon emission off
a muon.  In the absence of photon recombination, the lepton masses
serve as a physical cutoff for collinear singularities.  On the
computational side, this simply forbids the recombination of a muon
with momentum $p_\mu$ and a photon with momentum $p_\gamma$ to a
quasi-particle of momentum $p_{\mu\gamma}=p_\mu+p_\gamma$ in the
collinear regions as one would do in a collinear-safe setup.  In the
case of photon emission off electrons, the detection of the two
particles takes place in the electromagnetic calorimeter. The finite
resolution of the detector then defines a natural ``cone size'' for
the recombination of the lepton--photon pair to a single
quasi-particle.  In our collinear-unsafe setup, we exclude the muons
from recombination, while the electrons are recombined with photons
like in the collinear-safe case.

In \citere{Dittmaier:2008md}, the dipole subtraction formalism was
extended to collinear-unsafe observables.  As in the collinear-safe
case, the starting point of the formalism is
Eq.~\refeq{dNLO-masterFormula}, the fundamental difference being that
without recombination of a lepton--photon pair, some observables may
now be sensitive to the individual lepton and photon momenta within
the collinear region.  While this is obvious for the real-emission
matrix element, this is also required from the subtraction terms in
order to guarantee the local subtraction of the singularities.  To
this end, the reduced $n$-point kinematics of the local subtraction
terms (which are integrated over the $(n+1)$-particle phase space) is
a posteriori extended to an effective $(n+1)$-particle configuration
with a resolved collinear lepton--photon pair with momenta
\begin{align}
 \tilde p_1,\ldots,p_i=z\tilde p_i,\ldots,\tilde p_n,p_\gamma=(1-z)\tilde p_i.
\end{align}
Here $\tilde p_i$ denotes the momentum of a particular final-state
emitter in the reduced kinematics, and $p_i$ its momentum after
collinear $\ga$ emission.  The energy fraction of the emitter in the
lepton--photon pair is denoted by $z$; 
it is constructed from
kinematical invariants of the $(n+1)$-particle phase space.  The local
subtraction terms are evaluated in the same way as in the
collinear-safe case. However, any collinear-unsafe contribution is now
cut with respect to the unrecombined $(n+1)$-particle phase space:
\begin{align}
 &\Delta\sigma_{\rm NLO} =\int_{n+1} {\rm d}\Phi\Big[|{\cal{M}}_{\rm real}|^2\Theta(p_1,\ldots,p_n,p_\gamma)-|{\cal{M}}_{\rm sub}|^2\Theta(\tilde p_1,\ldots,\tilde p_n;z)\Big]\label{dNLOqqfinite-colunsafe}\\
 &+ \int_{n}{\rm d}\tilde\Phi\int_0^1{\rm d}x\int_0^1{\rm d}z\left[|{{\cal{M}}_{\rm sub}(x,z)}|^2_{\rm fin} + \delta(1-x)\delta(1-z)2{\rm Re}\left[{{\cal{M}}^*_{\rm Born}{\cal{M}}_{\rm virt}}\right]_{\rm fin}\right]\Theta(\tilde p_1,\ldots,\tilde p_n;z).\nonumber
\end{align}
The schematic shorthand notation 
\begin{align}
\Theta(\tilde p_1,\ldots,\tilde p_n;z)\equiv\Theta(\tilde p_1,\ldots,p_i=z\tilde p_i,\ldots,\tilde p_n,p_\gamma=(1-z)\tilde p_i)
\end{align}
applies separately to every dipole with final-state emitter momentum
$p_i=z \tilde p_i$ after $\ga$ emission. Note that the one-particle
phase-space integral $[{\rm d}k]$ in the integrated subtraction terms
in Eq.~\refeq{dNLO-masterFormula} is modified, because the $z$
dependence is required in the re-added subtraction contribution in
order to allow for cuts on the bare lepton momentum also in the
collinear region. This is in contrast to the collinear-safe case where
$z$ could be integrated in a process-independent way
\cite{Dittmaier:1999mb}. Splitting off the soft- and
collinear-singular contributions in the $z$-integration, the
subtraction terms can be separated into an inclusive part for the
collinear-safe case plus extra terms for the collinear-unsafe case.
The detailed form of $|{{\cal{M}}_{\rm sub}(x,z)}|^2_{\rm fin}$ can be
found in \citere{Dittmaier:2008md} and is not repeated here.

For collinear-unsafe observables, the integration over $z$ is not
inclusive, so that the conditions for the KLN theorem are not
fulfilled. Hence, the collinear singularities from the virtual
corrections do not entirely cancel against those from the real
corrections. Using the muon mass $m_\mu$ as a physical regulator,
terms of order $\alpha\ln(m_\mu)$ remain in the integral and modify
the cross section, which often leads to significant shape distortions
of differential distributions. Since partonic scattering energies at
the LHC are much larger than the muon mass, all terms suppressed by
factors of $m_\mu$ can be safely neglected. From a practical point of
view this means that all kinematics is evaluated with exactly massless
muons and the relicts from collinearly-sensitive observables remain in
the finite but possibly large contributions of order
$\alpha\ln(m_\mu)$.

\begin{figure}
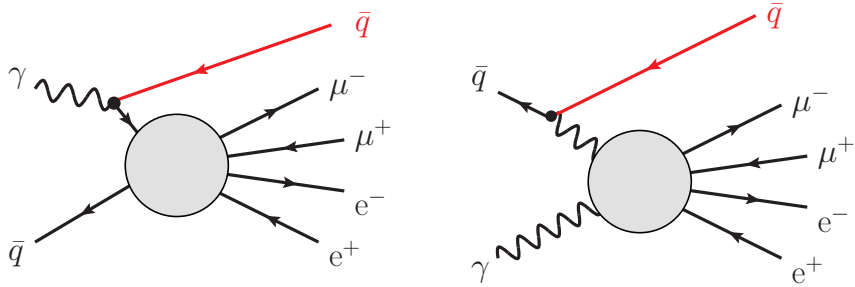

  \begin{center}
    \begin{minipage}{0.35\textwidth}
      \includegraphics[width=\textwidth]{diagrams/{{RealphQQ}}}
    \end{minipage}
    \quad
    \begin{minipage}{0.35\textwidth}
      \vspace{1mm}
      \includegraphics[width=\textwidth]{diagrams/{{RealQphQ}}}
    \end{minipage}
  \end{center}
  \caption{Photon-induced contributions with the two initial-state
    splittings $\gamma \to \bar q+q^*$ and $\bar q \to \bar q +
    \gamma^\star$. The star indicates the particle belonging to the 
    initial state of the reduced Born matrix element.}
\label{fig:realPhotonInduced}
\end{figure}

\paragraph{Quark--photon-induced channels:} 
The $q\gamma$-induced real contributions include all channels with one
photon and one (anti)quark in the initial state. Since an external
soft quark does not lead to a singularity and since there is no
collinear divergence when the final-state quark becomes collinear to
one of the final-state leptons, the matrix elements exhibit only
collinear initial-state singularities. As illustrated in
\reffi{fig:realPhotonInduced}, they can be grouped into two classes:
First, the incoming photon splits into a quark--antiquark pair with
the final-state (anti)quark becoming collinear to the incoming photon,
or second, the incoming (anti)quark splits into a
photon--\mbox{(anti)quark} pair with the final- and initial-state
(anti)quark becoming collinear. With the dipole subtraction method
each of the two collinear singularities may be locally subtracted with
a single dipole whose functional form is given in
\citere{Dittmaier:2008md}. The singularities in the two corresponding
integrated subtraction terms cancel against the collinear counterterm
from the PDFs, which can, \eg, be found in \citere{Dittmaier:2009cr}.

\subsection{Numerical implementation and independent checks of the calculation}
We have performed a complete calculation of all contributions using
the publicly available matrix element generator
{\recola}~\cite{Actis:2016mpe} for the evaluation of the virtual
corrections and for all tree-level amplitudes at Born level and at the
level of the real corrections.  The phase-space integration has been
carried out with a multi-channel Monte Carlo integrator with an
implementation of the dipole-subtraction formalism
\cite{Catani:1996vz,Dittmaier:1999mb,Catani:2002hc,Dittmaier:2008md}
for collinear-safe and collinear-unsafe observables.  The calculation
has been cross-checked both at the level of phase-space points and
differential cross sections with two other independent
implementations.  The one-loop matrix elements of the equal-flavour
case have been checked against amplitudes from the {\sc Mathematica}
package {\sc Pole}~\cite{Accomando:2005ra}, which employs {\sc
  FeynArts}~\cite{Kublbeck:1990xc,Hahn:2000kx} and {\sc
  FormCalc}~\cite{Hahn:1998yk}.
The one-loop matrix elements of the mixed-flavour case have been
checked against a calculation based on diagrammatic methods like those
developed for four-fermion production in electron--positron collisions
\cite{Denner:2005es,Denner:2005fg}, starting from a generation of
amplitudes with {\sc FeynArts}~\cite{Kublbeck:1990xc,Hahn:2000kx} and
further algebraic processing with in-house {\sc Mathematica} routines.
In all three calculational approaches, the one-loop integrals are
evaluated with the tensor-integral library
\collier~\cite{Denner:2016kdg} containing two independent
implementations of the tensor and scalar integrals. \collier\ employs
the numerical reduction schemes of
\citeres{Denner:2002ii,Denner:2005nn} for one-loop tensor integrals
and the explicit results of one-loop scalar integrals of
\citeres{Beenakker:1990jr,Denner:1991qq,Denner:2010tr} for complex
masses.  The phase-space integration in all three approaches is
carried out with independent multi-channel Monte Carlo integrators
which are further developments of the ones described in
\citeres{Berends:1994pv,Dittmaier:2002ap,Motz:Thesis}.  For all
differential and total cross sections obtained with the different
implementations we find agreement within the statistical uncertainty
of the Monte Carlo integration.

\section{Phenomenological results}
\label{se:results}

\subsection{Input parameters}

In the numerical analysis presented below, we consider the LHC at a 
centre-of-mass (CM) energy of $13\TeV$ and choose the following input parameters.
For the values of the on-shell masses and widths of the gauge bosons we use
\begin{align}
  \qquad M_\PZ^{\rm os}&=91.1876 \GeV,& \Gamma_\PZ^{\rm os}&= 2.4952\GeV,&\nonumber\\
  \qquad M_\PW^{\rm os}&=80.385 \GeV,&  \Gamma_\PW^{\rm os}&=2.085\GeV.&
\end{align}
In the complex-mass scheme, the on-shell masses and widths 
need to be converted to pole quantities according to the relations~\cite{Bardin:1988xt}
\begin{align}
 M_V= \frac{M^{\rm os}_V}{\sqrt{1+(\Gamma^{\rm os}_V/M_V^{\rm os})^2 } }, \qquad\Gamma_V = \frac{\Gamma_V^{\rm os}}{\sqrt{1+(\Gamma_V^{\rm os}/M_V^{\rm os})^2 } }, \qquad V=\PW,\PZ.
\end{align}
The complex weak mixing angle used in the complex-mass scheme is
derived from the ratio $\mu_\PW/\mu_\PZ$, where $\mu_V^2=M_V^2-\ri
M_V\Gamma_V$.  For details on the complex renormalization of the EW
parameters, we refer to \citere{Denner:2005fg}.  Since the Higgs boson
and the top quark do not appear as internal resonances in our
calculation, their widths are set equal to zero.  For the
corresponding masses we choose the values
\begin{align}
  M_\PH&=125 \GeV,\qquad m_{\Pt}=173 \GeV.
\end{align}
All the charged leptons $\ell=\Pe,\mu,\tau$ and the quarks
$q=\Pu,\Pd,\Pc,\Ps,\Pb$ are considered as light particles with zero
mass throughout the calculation.  In the computation of
collinear-unsafe observables, the physical muon mass appears as a
regulator with numerical value
\begin{equation}
  m_{\Pmu}= 105.6583715 \MeV.
\end{equation} 
Note that this non-zero value for the muon mass is only kept in otherwise 
divergent logarithms from photon radiation off muons, while everywhere else 
the muons are strictly treated as massless particles.

We work in the $G_\mu$ scheme where the electromagnetic
coupling $\alpha$ is derived from the Fermi constant
\begin{equation}
 G_\mu=1.16637\times 10^{-5} \GeV^{-2}
\end{equation}
according to
\begin{align}
 \alpha_{G_\mu}=\frac{\sqrt{2}}{\pi}G_\mu 
 M_\PW^2\left(1-\frac{M_\PW^2}{M_\PZ^2}\right).
\end{align}
This choice absorbs the effect of the running of $\alpha$ from
zero-momentum transfer to the EW scale into the LO cross section and
thus avoids mass singularities in the charge renormalization.
Moreover, $\alpha_{G_\mu}$ partially accounts for the leading
universal renormalization effects originating from the
$\rho$-parameter. The fine-structure constant,
\begin{equation}
 \alpha(0)=1/137.035999679
\end{equation}
is only used as coupling parameter in the relative photonic
corrections, \ie the NLO contribution $\Delta\sigma_{\rm NLO}^{\rm
  phot}$ to the cross section scales as $\alpha_{G_\mu}^4\alpha(0)$.
This choice is motivated by the dominance of corrections from the
emission of real photons coupling with $\alpha(0)$.  The relative
genuine weak corrections, on the other hand, are parametrized with
$\alpha_{G_\mu}$, \ie $\Delta\sigma_{\rm NLO}^{\rm weak}$ scales as
$\alpha_{G_\mu}^5$, since the dominating weak high-energy corrections
involve this coupling factor.  As we do not consider QCD corrections
in this paper, our cross-section predictions do not depend on the
renormalization scale $\mu_{\rm ren}$.  The dependence of the relative
NLO EW corrections on the factorization scale $\mu_{\rm fact}$ is only
marginal, so that we simply set it equal to the Z-boson mass,
$\mu_{\rm fact}=M_\PZ$, without the need to investigate residual scale
dependences or alternative scale choices.  As PDFs we use the {\tt
  NNPDF23\_nlo\_as\_0118\_qed} set \cite{Ball:2013hta}.%
\footnote{In this calculation we take the photon density of
{\tt NNPDF23\_nlo\_as\_0118\_qed} in spite of its larger uncertainty
compared to the LUXqed photon distribution~\cite{Manohar:2016nzj},
in order to rely on one consistent PDF set. This procedure is further justified
by the fact that $\gamma\gamma$ and $q\gamma$ contributions turn out to be 
phenomenologically negligible.}
Throughout our calculation of EW corrections, we employ the deep-inelastic-scattering 
(DIS) factorization scheme, following the arguments given in \citere{Diener:2005me}.
The corresponding finite terms for the EW corrections to be included in the 
subtraction formalism can be found in \citere{Dittmaier:2009cr}.

\subsection{Definition of observables and acceptance cuts}
\label{sec:observablesAndCuts}

In the following we define two different event selections: an
``inclusive'' and a ``Higgs-specific'' setup.  The former uses typical
lepton-identification cuts without any further selection criteria; the
latter is motivated by specific criteria designed for Higgs-boson
analyses by ATLAS~\cite{Aad:2014eva} and
CMS~\cite{Chatrchyan:2013mxa}.

\paragraph{Inclusive setup:}
In the collinear-safe case, photons emerging in the real-emission
contributions are recombined with the closest charged lepton (\cf
\refse{sec:realEmission}) if their separation in the
rapidity--azimuthal-angle plane obeys
\begin{equation}
\label{eq:recombi}
 \Delta R_{\ell_i,\gamma} = \sqrt{(y_{\ell_i}-y_\gamma)^2+(\Delta\phi_{\ell_i\gamma})^2} < 0.2.
\end{equation}
Here, 
$y_j$ 
denotes the rapidity of a final state particle $j$ and $\Delta\phi_{\ell_i\gamma}$ the 
azimuthal-angle difference between a charged lepton $\ell_i$ and the photon $\gamma$.
Note that we only take into account photons with $|y_\gamma|<5$ in the recombination procedure, while we
consider photons with larger rapidities as lost in the beam pipe.
In the collinear-unsafe case, photons are recombined only with electrons/positrons, while no recombination with 
muons/antimuons is performed. 
For observables of the equal-flavour-lepton final state,
the leading lepton pair $(\ell_1^+,\ell_1^-)$ is defined as the one
whose invariant mass is closest to the nominal Z-boson mass; the
subleading lepton pair $(\ell_2^+,\ell_2^-)$ is then formed by the remaining two leptons.
Leading leptons are thus labelled as $\ell_1^\pm$,  
and subleading leptons as $\ell_2^\pm$.

As default setup, we consider a minimal set of selection cuts inspired
by the ATLAS analysis~\cite{Aad:2014wra}.
For each charged lepton $\ell_i$, we restrict transverse momentum $p_{{\rm T},\ell_i}$ and 
rapidity $y_{\ell_i}$ according to
\begin{align}
 p_{{\rm T},\ell_i}>p_{\rm T,min}=15\GeV,\qquad |y_{\ell_i}|&<2.5.
 \label{eq:inc-cuts1}
\end{align}
Any pair of charged leptons $(\ell_i,\ell_j)$ is required to be well separated in the rapidity--azimuthal-angle plane, \begin{equation}
\label{eq:inc-cuts2}
 \Delta R_{\ell_i,\ell_j} >0.2.
\end{equation}
\paragraph{Higgs-specific setup:}
For the Higgs-specific setup, 
motivated by the ATLAS and CMS
analyses~\cite{Aad:2014eva,Chatrchyan:2013mxa}, 
we replace the cuts of Eq.~\refeq{eq:inc-cuts1} by
the less restrictive criteria
\begin{align}
 \label{eq:hcuts3}
 p_{{\rm T},\ell_i}>p_{{\rm T},{\rm min}}'=6\GeV,\qquad |y_{\ell_i}|&<2.5,
\end{align}
retain the cut \refeq{eq:inc-cuts2},
and complement them 
with additional invariant-mass cuts on the charged leptons. For the mixed-flavour final state, 
we require for the two same-flavour lepton pairs 
\begin{align}
\label{eq:hcuts1}
  40\GeV<\,& M_{\ell_1^+,\ell_1^-}\,<120\GeV,\nonumber\\
  12\GeV<\,& M_{\ell_2^+,\ell_2^-}\,<120\GeV,
\end{align}
with $M_{\ell_1^+,\ell_1^-}$ and $M_{\ell_2^{+},\ell_2^{-}}$ referring
to the invariant masses of the lepton pair that is closer to or
further away from the nominal Z-boson mass, respectively. For the
same-flavour final state, we apply the cuts of Eq.~\refeq{eq:hcuts1}
after selecting the leading and subleading lepton pairs
$(\ell_1^+,\ell_1^-)$ and $(\ell_2^+,\ell_2^-)$ in the same way as
described above.  The invariant mass
$M_{4\ell}$ of the four-lepton system is subjected to the cut
\begin{align}
\label{eq:hcuts2}
  M_{4\ell}>100\GeV,
\end{align}
which is independent of the flavour of the final-state leptons.

In both setups we treat the additional (anti)quark in the final state
of the photon-induced contributions in a fully inclusive way, \ie we
do not apply any jet veto.  Finally, we note that the employed
single-particle lepton identification cuts are chosen to be equal for
all charged leptons.  Since the lepton pairing in the equal-flavour
final state is flavour independent, the results for the
$\Pe^+\Pe^-\Pe^+\Pe^-$ final state are equal to the results of the
$\mmmm$ final state within the collinear-safe setup.  In the
following, $[4\mu]$ denotes the equal-flavour final state, while
$[2\mu2\Pe]$ denotes the mixed-flavour final state.

Whenever possible, we have compared the results of our full off-shell
calculation with the available results for on-shell Z-boson pair
production \cite{Bierweiler:2013dja,Baglio:2013toa}. Since our
calculation sets phase-space cuts on the charged leptons, a direct
comparison is not possible for most observables.  Moreover, the
calculations for stable Z~bosons do not take into account corrections
to the Z-boson decays.  Finally, there are differences in the setup:
The values in the literature are given for a CM energy of
$\sqrt{s}=14\TeV$, different factorization scales and PDFs.
Nevertheless, the relative EW corrections to those observables that
are less sensitive to off-shell effects and corrections to the Z-boson
decays can still be directly compared. This holds in particular for
the Sudakov enhancement at large transverse momentum of the on-shell
Z~boson or the corresponding charged final-state lepton pair.

\subsection{Results on integrated cross sections}
\label{ssec:results-xsec}
The results for the integrated cross sections of the processes
$\ppmmee$ and $\ppmmmm$ at a CM energy of $13 \TeV$ are summarized in
\refta{tab:inc-xsec}. Results are given for the inclusive selection
cuts of Eqs.~\refeq{eq:inc-cuts1}--\refeq{eq:inc-cuts2} and the
Higgs-specific selection cuts of
Eqs.~\refeq{eq:inc-cuts2}--\refeq{eq:hcuts2}.  Together with the LO
cross sections, the NLO EW corrections to the ${\bar q q}$
contribution and the relative contributions of the photon-induced
channels $\delta_{\gamma\gamma}=\sigma_{\gamma\gamma}/\sigma_{\bar q
  q}^\mr{LO}$ and $\delta_{q\gamma}=\sigma_{q\gamma}/\sigma_{\bar q
  q}^\mr{LO}$ are shown.  The ${\bar q q}$ contribution
$\delta^\mr{EW}_{\bar q q}=\Delta\sigma_{\bar q
  q}^\mr{EW}/\sigma_{\bar q q}^\mr{LO}$ is split into the purely weak
and the photonic part, $\delta^\mr{weak}_{\bar q q}$ and $\delta_{\bar
  q q}^\mr{phot}$, respectively, so that $\delta^\mr{EW}_{\bar q
  q}=\delta^\mr{weak}_{\bar q q}+\delta_{\bar q q}^\mr{phot}$.  For
the photonic corrections, we further distinguish between the
collinear-safe setup $\delta_{\bar q q}^\mr{phot,safe}$, where the
bremsstrahlung photon is recombined with any charged lepton, and the
collinear-unsafe case $\delta^\mr{phot,unsafe}_{\bar q q}$, where the
muons are excluded from recombination, as described in
\refse{sec:realEmission}.  Since we define $\sigma_{\bar q q}^\mr{LO}$
from NLO PDFs (instead of LO PDFs), the resulting relative corrections
$\delta_{\bar q q}^\mr{EW/weak/phot}$ to the $\bar qq$ channels are
practically independent of the chosen PDF set, rendering our results
for $\delta_{\bar q q}^\mr{EW/weak/phot}$ well suited for a
combination with state-of-the-art QCD predictions.

\begin{table}
\begin{center}
\begin{tabular}
{|l|c|cccccc|}
\hline\vspace{-4mm}
&&&&&&&\\
&$\sigma_{\bar q q}^\mr{LO}$~[fb] &  $\delta^\mr{weak}_{\bar q q}(\%)$&  $\delta_{\bar q q}^\mr{phot,safe}(\%)$& $\delta^\mr{phot,unsafe}_{\bar q q}(\%)$ & $\delta_{\gamma\gamma}(\%)$ &$\delta_{q \gamma}(\%)$ & 
\\
\hline
\hline
%\\
% incl. [2$\mu$2e] & 11.4962(4) & $-4.32$ & $-0.93$ & $-1.68$ & $+0.1284$  & $+0.017$ & 
 incl. [2$\mu$2e] & 11.4962(4) & $-4.32$ & $-0.93$ & $-1.68$ & $+0.13$  & $+0.02$ & 
\\
\hline
% incl. [4$\mu$]  &  \phantom{1}5.7308(3) & $-4.32$ & $-0.94$ & $-2.43$ & $+0.1145$ & $+0.019$ &
 incl. [4$\mu$]  &  \phantom{1}5.7308(3) & $-4.32$ & $-0.94$ & $-2.43$ & $+0.11$ & $+0.02$ &
\\
\hline
% Higgs [2$\mu$2e]& 13.8598(3) & $-3.59$ & $-0.04$ & $-0.28$ & $+0.229$ &  $-0.09$ & 
 Higgs [2$\mu$2e]& 13.8598(3) & $-3.59$ & $-0.04$ & $-0.28$ & $+0.23$ &  $-0.09$ & 
\\
\hline
% Higgs [4$\mu$]  &  \phantom{1}7.1229(2) & $-3.42$ & $-0.09$ & $-0.66$ & $+0.3042$ & $-0.14$ &
 Higgs [4$\mu$]  &  \phantom{1}7.1229(2) & $-3.42$ & $-0.09$ & $-0.66$ & $+0.30$ & $-0.14$ &
\\
\hline
\end{tabular}
\end{center}
\caption{LO cross sections for $\ppmmee$ and $\ppmmmm$ with the relative corrections $\delta_i =\sigma_i/\sigma_{\bar q q}^\mr{LO}$ 
for the LHC at $\sqrt{s}=13\TeV$. Results are given for the inclusive
setup (``incl.") with the cuts of
Eqs.~(\ref{eq:inc-cuts1})--(\ref{eq:inc-cuts2}) and the Higgs-specific
setup (``Higgs") with the selection cuts of
Eqs.~(\ref{eq:inc-cuts2})--(\ref{eq:hcuts2}), respectively.} 
\label{tab:inc-xsec}
\end{table}

We first analyse the inclusive scenario.  The major contribution to
the corrections stems from the $\bar qq$~channels where the purely
weak correction of $-4.3\%$ has the highest impact.  The
photon-induced contribution matters only at the permille level,
justifying that we neglect higher-order corrections to this channel.
Besides the small photon flux in the proton, yet another reason for
the strong suppression is that channels with two photons in the
initial state involve at most one resonant Z~boson, as illustrated by
the sample diagram of \reffi{fig:born}(c). We count such kinematical
topologies as background topologies to the dominant contribution with
two possibly resonant Z-boson propagators as they appear in the ${\bar
  q q}$
channels.% 
\footnote{Note that in W-pair production also the $\gamma\gamma$
  channel has an enhanced signal topology with two W-boson resonances
  due to the coupling of the W~boson to the photon.}  The
$q\gamma$-induced corrections $\delta_{q \gamma}$ are even further
suppressed by yet another order of magnitude and are thus entirely
negligible in the integrated cross section.

Summing up all contributions, we find for both the $[2\mu2\Pe]$ and
the $[4\mu]$ final state the same correction of $-5.1\%$ in the
collinear-safe setup.  The corrections to vector-boson pair production
with on-shell Z~bosons are known to be $\sim-4.5\%$
\cite{Bierweiler:2013dja}, respectively $\sim-4\%$
\cite{Baglio:2013toa} in a slightly different setup. The differences
can be attributed to the off-shell effects, including also additional
virtual photon exchange, and to differences in phase-space cuts and in
the employed numerical setup (\cf comments at the end of
\refse{sec:observablesAndCuts}). Note that there is no photon-induced
contribution of the form $\gamma\gamma\to\PZ\PZ$ with on-shell
Z~bosons in the final state. The strong suppression of the $q\gamma$
channels confirms similar findings in the on-shell approximation
\cite{Baglio:2013toa}.

Comparing the collinear-unsafe photonic corrections with the
collinear-safe case, we observe differences in $\delta_{\bar q
  q}^\mr{phot}$ by $\sim0.7\%$ and $\sim1.5\%$ for $[2\mu2\Pe]$ and
$[4\mu]$ final states, respectively. In the collinear-unsafe case,
final-state radiation off muons is enhanced through the mass
singularity $\propto\alpha\ln(m_\Pmu)$ in the phase-space integral.
Since the photon is not recombined with the muons, there are
systematically more events with a large energy loss in one of the muon
momenta induced by final-state radiation. For this reason, less events
pass the event-selection in the collinear-unsafe case, leading to more
negative corrections. This ``acceptance correction" scales with the
number of leptons treated in a collinear-unsafe way, explaining the
factor of two between the $[2\mu2\Pe]$ and the $[4\mu]$ cases in
$\delta^{\rm safe}-\delta^{\rm unsafe}$.

The amplitudes for the equal-flavour final state can be obtained from
the different-flavour amplitudes by antisymmetrization with respect to
a pair of equal final-state leptons. At the level of squared matrix
elements, diagrams with a $\mu^+\mu^-$ pair originating from a single
vector boson interfere with diagrams where the $\mu^+$ and the $\mu^-$
originate from two different vector bosons. These interference terms
lead to a deviation from a naive rescaling of the different-flavour
cross section by a symmetry factor of two. From the ratio $\sigma_{\rm
  LO}[2\mu2\Pe]/(2\sigma_{\rm LO}[4\mu])\approx 1.003$ we find a
negative interference of about 0.3\% for the integrated LO cross
section. Comparing this number with the total relative correction of
$-5.1\%$ for both final states, we conclude that the impact of
interferences on the total cross section in the inclusive setup is a
``LO effect'' in the sense that the relative corrections do not modify
this behaviour.  The reason for the smallness of the interferences is
that the LO cross section is dominated by contributions with two
resonant Z-boson propagators. In the interference terms, however, in
at least one diagram both Z-boson propagators are off shell.  This
explains also why the impact of the interferences in the
$\gamma\gamma$~Born cross section alone is less suppressed with
$\sigma_{\gamma\gamma}[2\mu2\Pe]/(2\sigma_{\gamma\gamma}[4\mu])\approx
1.12$ since this is a background contribution with at most one
resonant Z~boson.

We now turn to the Higgs-specific setup.  Despite the additional cuts
of Eqs.~(\ref{eq:hcuts1})--(\ref{eq:hcuts2}), the cross sections for
this scenario are larger than for the previously considered inclusive
setup.  This feature is due to the less severe cut of $6\GeV$ imposed
on the transverse momenta of the charged leptons in the Higgs-specific
setup, as compared to a cut of $15\GeV$ in the inclusive setup.
Moreover, we observe a percent-level deviation at LO from the naive
scaling factor of two between the equal- and unequal-flavour cases,
$\sigma_{\rm LO}[2\mu2\Pe]/(2\sigma_{\rm LO}[4\mu])\approx 0.973$.
This reflects, on the one hand, an expected enhancement of the
interference terms since, by construction, the whole scenario is more
sensitive to the off-shell effects of the vector bosons. On the other
hand, the additional invariant-mass cuts in Eq.~\refeq{eq:hcuts1}
depend in the equal-flavour case on the chosen lepton-pairing
algorithm, while they do not in the mixed-flavour case. A quantitative
statement on the size of the pure interference effects would thus only
be possible if the same lepton pairing was applied also in the
mixed-flavour case. The same arguments apply also to the corrections
in \refta{tab:inc-xsec}, \ie the differences between the equal-flavour
and the mixed-flavour final state are due to interferences and event
selection. As a general pattern, we observe that the bulk of
corrections to the total cross sections of around $-3.5\%$ stems from
the weak corrections, while photonic corrections contribute only at
the sub-percent level.

\subsection{Results on differential cross sections in the inclusive
  setup}\label{se:diff-results-inclusive} 

{\it Invariant-mass and transverse-momentum distributions} 
\\[.5em]
In order to illustrate the impact of EW corrections on differential
observables, we present several results 
on distributions in the inclusive setup at a proton--proton CM energy
of $\sqrt{s}=13\TeV$.  We choose the collinear-safe setup as
default and provide selected results within the collinear-unsafe case
subsequently.

Figure~\ref{fig:inc-m4lep} shows the four-lepton invariant-mass 
distribution for the unequal-flavour $[2\mu2\Pe]$ and the
equal-flavour $[4\mu]$ final states.  The left-hand side resolves the
off-shell region with its threshold and resonance structures 
with a fine histogram binning, while the panels on the right-hand side
show the whole range from the off-shell region over the resonances and
thresholds up to the TeV regime in coarse-grained resolution.  The
absolute predictions of the LO and NLO distributions of both the
$[2\mu2\Pe]$ and $[4\mu]$ final states in the upper panels follow 
the characteristic pattern of Z-boson pair production: The first peak
around $M_{4\ell}=M_\PZ$ represents the single-resonant production of
the Z~boson in the $s$-channel according to the sample diagram in
\reffi{fig:born}(b), the threshold at $M_{4\ell}=2M_\PZ$ stems from 
doubly-resonant diagrams 
as depicted in the diagram in \reffi{fig:born}(a). The knee 
above $M_{4\ell}=M_\PZ+2p_{{\rm T},{\rm min}}\approx120\GeV$ is
induced by the kinematical cut of Eq.~\refeq{eq:inc-cuts1} on the lepton
transverse momentum. For $M_{4\ell}\gtrsim M_\PZ+2p_{{\rm T},{\rm
    min}}$, the cross section is dominated by events with one resonant
Z~boson $(\to \ell_1^+\ell_1^-)$ and the $\ell_2^+\ell_2^-$ pair with
$M_{\ell_2^+\ell_2^-}\gtrsim 30\GeV$, where both 
lepton pairs
are almost at rest in the transverse plane. Since
$E_{\ell^+_2}+E_{\ell^-_2}>2p_{{\rm T},{\rm min}}= 30\GeV$ is necessary for
the event to pass the cuts, 
$M_{4\ell}< M_\PZ+2p_{{\rm T},{\rm min}}$ 
is only possible if $M_{\ell_1^+\ell_1^-}\le E_{\ell_1^+}+E_{\ell_1^-}< M_\PZ$ which
drastically reduces the transition rate, since no resonance
enhancement is present anymore in the diagram type illustrated in 
\reffi{fig:born}(a).

\bfi
\begin{center}
\begin{minipage}{0.49\textwidth}
 \includegraphics[width=\textwidth]{finalplots/{{ATLASincl.13TeV.invmass.4lep.ratio.4mu.2mu2e.highres.collsafe}}}
\end{minipage}
\begin{minipage}{0.49\textwidth}
  \includegraphics[width=\textwidth]{finalplots/{{ATLASincl.13TeV.invmass.4lep.ratio.4mu.2mu2e.lowres.collsafe}}}
\end{minipage}
\end{center}
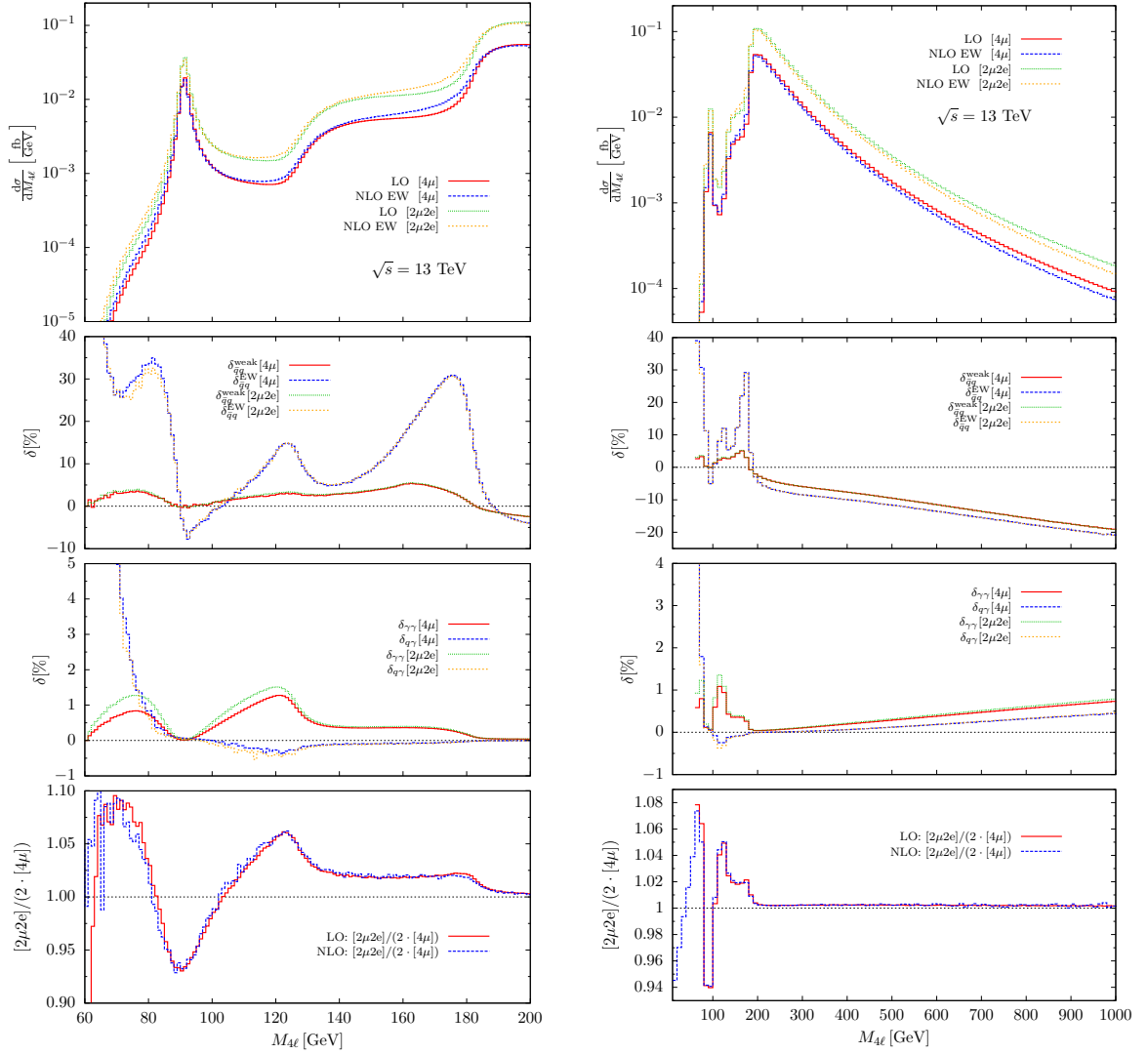
\caption{Invariant-mass distribution of the four-lepton system (upper panels),
  corresponding EW corrections (2nd panels from above), $\gamma\gamma$ and $q\gamma$
  contributions (third panels from above) for the unequal-flavour
  $[2\mu2\Pe]$ and the equal-flavour $[4\mu]$ final states in the
  inclusive setup. The panels at the bottom show the ratio of the $[2\mu2\Pe]$
  and $[4\mu]$ final states.}
\label{fig:inc-m4lep}
\efi

The panels directly below the absolute predictions for the cross
sections show the relative EW corrections to the $\bar qq$ channels in
the collinear-safe setup, comparing the purely weak contribution with
the full EW contribution (EW=weak+phot). Apart from the off-shell
region below the single Z~resonance, we observe that the relative EW
corrections of the mixed-flavour final state and the equal-flavour
final state are equal over the whole invariant-mass spectrum. This
confirms at the level of differential distributions that the
interference effect is mainly a LO effect, in accordance with
what we have already seen for the integrated cross section. 

\begin{sloppy}
  The four-lepton invariant mass in the inclusive setup is well suited
  to study the relative size of the interferences, as this observable
  does not depend on the lepton pairing.  We show in the lowest panels
  of \reffi{fig:inc-m4lep} and the following figures the ratio $({\rm
    d}\sigma_{\rm (N)LO}[2\mu2\Pe]/{\rm d}{\mathcal{O}})/(2{\rm
    d}\sigma_{\rm (N)LO}[4\mu]/{\rm d}{ \mathcal{O}})$, where
  $\mathcal{O}$ denotes the considered observable, \eg $M_{4\ell}$ in
  \reffi{fig:inc-m4lep}.  The LO and NLO curves are, as expected,
  almost equal.  The size of the interference effect varies in the
  region where no lepton pair is resonant from $-7\%$ at
  $M_{4\ell}=M_\PZ$ to $+6\%$ at $M_{4\ell}=M_\PZ+2p_{{\rm T},{\rm
      min}}$.  Thus, the unequal-flavour matrix elements cannot
  describe the equal-flavour final state there.  In the region
  $M_\PZ+2p_{{\rm T},{\rm min}}\lesssim M_{4\ell}\lesssim 2M_{\PZ}$,
  where only one lepton pair can be resonant, the interference effect
  amounts to $2\%$.  Above the ZZ threshold, the ratio is equal to one
  up to fractions of a percent, since in this region of phase space
  the doubly-resonant contribution dominates over any non-resonant
  interference effect.  For higher invariant masses $M_{4\ell}$, the
  overlap of the two resonance pairs becomes smaller and smaller in
  phase space, so that the ratio asymptotically approaches~one.
\end{sloppy}

We inspect the EW corrections in more detail. In the
high-invariant-mass region, the correction is entirely dominated by
the purely weak contribution and reaches about $-20\%$ around $1\TeV$.
The $M_{4\ell}$ distribution at high scales is not dominated by the
Sudakov regime of ZZ production where all Mandelstam variables
($s,t,u$) of the $2\to2$ particle process would have to be large.
Instead, Z-pair production at high energies is dominated by
forward/backward-produced Z~bosons, where $t$ and $u$ are small.  At
the ZZ production threshold, the weak corrections change their sign
and reach up to $5\%$ below.  Note that this non-trivial sign change
makes it impossible to approximate the full NLO EW results by a global
rescaling factor.  At the Z~peak, the weak corrections are extremely
suppressed.  The photonic corrections remain almost constant above the
ZZ threshold, at around $-2\%$ to $-3\%$, and show the typical
radiative tails: Below a threshold or close to a resonance, the LO
cross section falls off steeply. Final-state radiation of a real
photon, however, may shift the value of the measured invariant mass to
smaller values. Since the LO cross section is small in this
phase-space region, the relative correction due to the bremsstrahlung
photon becomes large. The structure of the radiative tails follows
precisely the thresholds and the resonances at leading order: one
below $2M_\PZ$, one near $M_\PZ+2p_{\rm T,{\rm min}}\approx120\GeV$,
and another one below the Z~resonance.

For completeness, we also show the photon-induced corrections in a
separate panel (third row in \reffi{fig:inc-m4lep}).  Over the whole
range of the distribution, both the $\gamma\gamma$ and $q\gamma$
contributions are strongly suppressed with respect to the $\bar
qq$~processes.  Since the $\gamma\gamma$ channel has only a single
Z~resonance according to the diagram in \reffi{fig:born}(c), it is
strongest suppressed with respect to the $\bar q q$~LO cross section
above the ZZ threshold and near the $s$-channel resonance at $M_\PZ$.
In the non-resonant region below $\MZ+2p_{\rm T,min}$ it reaches up to
$1\%$.  Since there the LO cross section is small anyway, the overall
impact remains small, in agreement with the result for the integrated
cross section. Differences between the equal-flavour and the
unequal-flavour final states due to interferences are only visible
below $M_\PZ+2p_{\rm T,{\rm min}}$ where none of the lepton pairs is
resonant.  The $q\gamma$ channels contribute over the whole spectrum
at most at the permille level. The large correction near the
phase-space boundary is phenomenologically irrelevant as the
corresponding LO cross section in this region is very small anyway.
Due to the negligible impact of any photon-induced corrections in the
inclusive setup, we do not show them separately any more in the
following plots.

Up to the details in the event selection and the corrections from the
Z-boson decays, the four-lepton invariant mass $M_{4\ell}$ can be
compared to the ZZ-invariant-mass distribution obtained in the NLO
calculations of \citeres{Bierweiler:2013dja,Baglio:2013toa} with
on-shell Z~bosons.  The relative corrections to the distribution in
the invariant mass of the Z-boson pair at $M_{\PZ\PZ}=700\GeV$ are
given as $-11\%$ and $-8\%$ in \citeres{Bierweiler:2013dja} and
\cite{Baglio:2013toa}, respectively, while we find for the four-lepton
invariant mass $M_{4\ell}=700\GeV$ a relative correction of $-15\%$.
We attribute the difference mainly to the final-state radiation off
muons missing in the calculation with stable Z~bosons.  This
collinearly enhanced contribution typically leads to negative
corrections at the level of some percent, induced by the radiative
loss in transverse momenta that can potentially shift events out of
acceptance.

\bfi
\begin{center}
\begin{minipage}{0.49\textwidth}
 \includegraphics[width=\textwidth]{finalplots/{{ATLASincl.13TeV.invmass.mu+mu-11.ratio.4mu.2mu2e.highres.collsafe}}}
\end{minipage}
\begin{minipage}{0.49\textwidth}
 \includegraphics[width=\textwidth]{finalplots/{{ATLASincl.13TeV.invmass.mu+mu-22.ratio.4mu.2mu2e.highres.collsafe}}}
\end{minipage}
\end{center}
\caption{Invariant $\mu^+\mu^-$-mass distribution (upper panels),
  corresponding EW corrections (middle panels), and ratio of the
  $[2\mu2\Pe]$ and $[4\mu]$ final states (lower panels) in the
  inclusive setup.  In the left column the equal-flavour case is
  binned with respect to the leading lepton pair, while the right
  column shows results for the subleading one.}
\label{fig:inc-m2lep-highres}
\efi
The $\mu^+\mu^-$ invariant mass $M_{\mu^+\mu^-}$ is an example of an
observable where the differential cross section in the equal-flavour
case directly depends on the employed lepton pairing, as can be seen
in \reffi{fig:inc-m2lep-highres}.  The left column compares the
$\mu^+\mu^-$ invariant mass of the $[2\mu2\Pe]$ final state with the
one of the leading $\mu^+\mu^-$ pair of the equal-flavour case, while
the right column shows the same comparison, with the subleading
$\mu^+\mu^-$ pair instead. Note that we compare two different
observables, since the unequal-flavour case is binned with respect to
the flavour but not with respect to the kinematic ordering described
in \refse{sec:observablesAndCuts}.  Although this precludes us from
drawing conclusions on the interference effects (as we did for the
four-lepton invariant mass above), one may nevertheless learn from
such a comparison to which extent the mixed-flavour case can be used
to describe the equal-flavour case. Obviously, the shape of the
subleading lepton pair in the equal-flavour case widely resembles the
corresponding observable of the unequal-flavour final state (though it
is not equal) while the pattern of the leading lepton pair is
fundamentally different. The different behaviour can be explained as
follows: The $\mu^+\mu^-$ invariant-mass distribution of the
mixed-flavour case receives the largest contribution when the
corresponding $\Pe^+\Pe^-$ pair is on the Z~resonance. The situation
is similar when binning the subleading $\Pmu^+\Pmu^-$ pair where the
corresponding leading lepton pair is always closer to the resonance
and, thus, takes over the role of the $\Pe^+\Pe^-$ pair in the
mixed-flavour case for the dominant contribution.  Since interference
effects are strongly suppressed, this explains the almost constant
ratio $({\rm d}\sigma[2\mu2\Pe]/{\rm d}M_{2\ell})/(2{\rm
  d}\sigma[4\mu]/{\rm d}M_{2\ell})\approx 0.5$ above the resonance in
the lowest panel in the right column of \reffi{fig:inc-m2lep-highres};
it uniformly extends to larger values of $M_{\mu^+\mu^-}$ as can be
seen in the left columns of \reffi{fig:inc-m2lep-lowres}.  For this
particular observable away from the Z~resonance, the lepton pairing
thus basically identifies the two $\mu^+\mu^-$ pairs and the symmetry
factor of $1/2$ disappears. The large difference near the Z resonance
(see lowest panel in the right column of
\reffi{fig:inc-m2lep-highres}) is due to the fact that the subleading
muon pair is always further away from the resonance than the leading
pair, in contrast to the mixed-flavour case where the invariant mass
of the $\Pe^+\Pe^-$ pair is independent of the $\mu^+\mu^-$ pair.

By contrast, when binning the leading lepton pair, the other lepton
pair is forced to be further off shell with respect to the Z~resonance
and, hence, the distribution falls off much steeper with the distance
from $M_\PZ$.  The steeper drop in the interval
$160\GeV<M_{\mu^+\mu^-}<175\GeV$ is due to the fact that for invariant
masses $M_{\mu^+\mu^-}^{\mathrm{lead}}>2M_\PZ$ of the leading
$\mu^+\mu^-$ pair the invariant mass of the subleading $\mu^+\mu^-$
pair cannot be below $\MZ$ any more.%
\footnote{For $M_{\mu^+\mu^-}^{\mathrm{sublead}}$ below and
  $M_{\mu^+\mu^-}^{\mathrm{lead}}$ above the Z~resonance, the
  categorization of leading and subleading $\mu^+\mu^-$ pairs implies
  $\MZ-M_{\mu^+\mu^-}^{\mathrm{sublead}} >
  M_{\mu^+\mu^-}^{\mathrm{lead}}-\MZ$, so that
  $M_{\mu^+\mu^-}^{\mathrm{sublead}}<0$ would be required for
  $M_{\mu^+\mu^-}^{\mathrm{lead}}>2\MZ$.}  The broad peak between
$35\GeV$ and $60\GeV$ stems from the single $s$-channel resonance of
the four-lepton invariant mass at $M_{4\ell}=M_\PZ$.  The resonance
condition $M_{4\ell}^2=\sum_{\ell < \ell^\prime}2p_\ell
p_{\ell^\prime}=M_\PZ^2$ requires $\max(2p_\ell
p_{\ell^\prime})>M_\PZ^2/6$ implying a threshold for the leading
lepton pair of $M^{\mathrm{lead}}_{\mu^+\mu^-}>M_\PZ/\sqrt{6}\approx
37 \GeV$ in agreement with \reffi{fig:inc-m2lep-highres}. On the other
hand, the Z~resonance in $M_{4\ell}$ contributes only for
$M_{\mu^+\mu^-}^{\mathrm{lead}} < M_\PZ-2p_{\rm T,min}\approx 61
\GeV$, since the transverse momentum of the subleading leptons cannot
be lower than $p_{\rm T,min}$.  The $s$-channel resonance gives also
rise to a small bump in the invariant-mass distribution of the
subleading lepton pair at somewhat smaller invariant masses (see
r.h.s.\ of \reffi{fig:inc-m2lep-highres}).  The increase of the
distributions towards $M_{\mu^+\mu^-}\to 0$ in the subleading and
mixed-flavour $\mu^+\mu^-$ invariant masses are due to the tail of the
photon pole.  The spectrum of the leading lepton pair is
phenomenologically irrelevant at high invariant masses since it is
heavily suppressed due to the lack of a resonance enhancement.
\bfi
\begin{center}
\begin{minipage}{0.49\textwidth}
\includegraphics[width=\textwidth]{finalplots/{{ATLASincl.13TeV.invmass.mu+mu-22.ratio.4mu.2mu2e.lowres.collsafe}}}
\end{minipage}
\begin{minipage}{0.49\textwidth}
 \includegraphics[width=\textwidth]{finalplots/{{ATLASincl.13TeV.pt.mu+mu-22.ratio.4mu.2mu2e.collsafe}}}
\end{minipage}
\end{center}
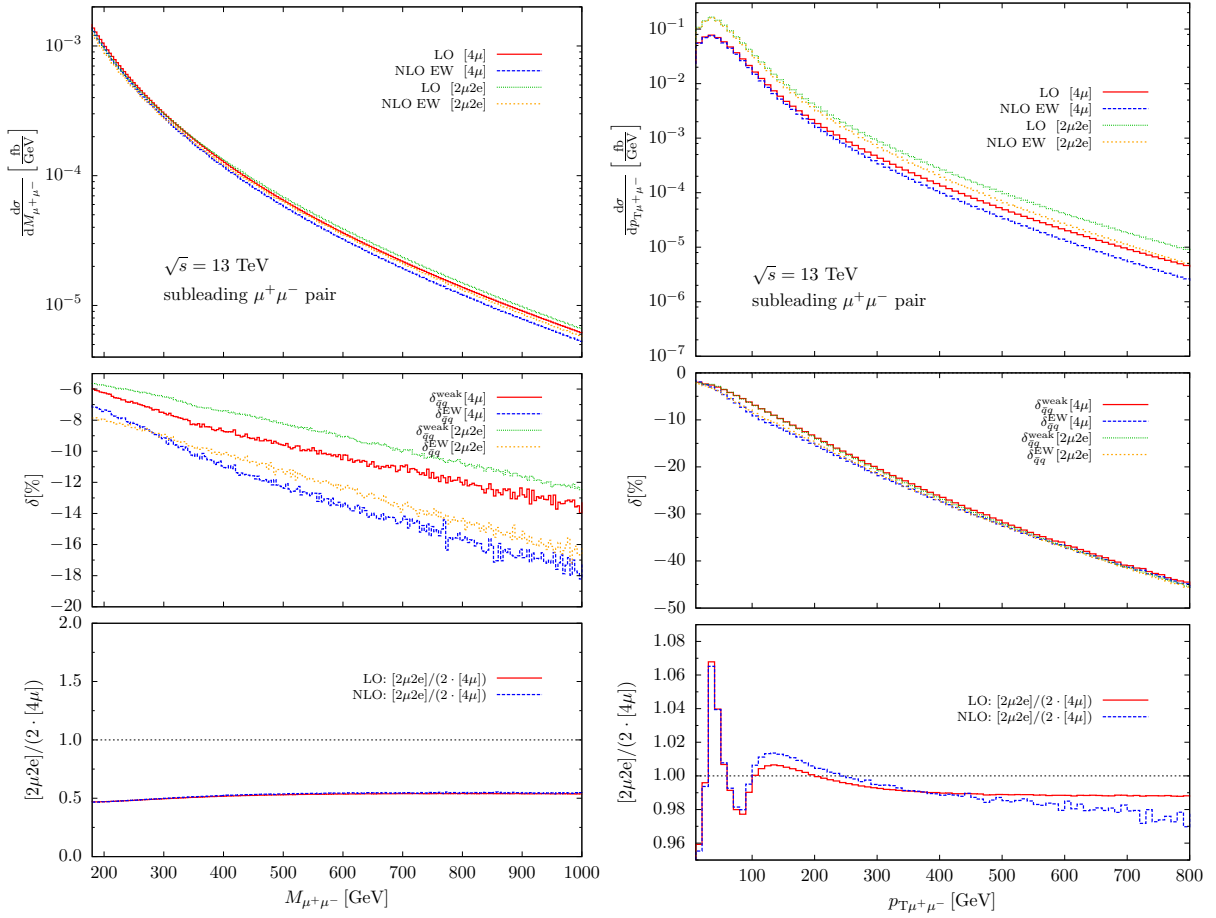
\caption{Invariant-mass (left) and transverse-momentum distribution
  (right) of the $\mu^+\mu^-$ pair (upper panels),
  corresponding EW corrections (middle panels), and ratio of the
  $[2\mu2\Pe]$ and $[4\mu]$ final states (lower panels) in the
  inclusive setup.  The equal-flavour case is binned with respect to
  the subleading lepton pair.}
\label{fig:inc-m2lep-lowres}
\efi

As discussed already in \citere{Biedermann:2016yvs} for the
$[2\mu2\Pe]$ final state (though for the Higgs-specific setup), the EW
corrections largely resemble the known structure of the photonic and
weak corrections to Drell--Yan-like single-Z production, which are,
e.g., discussed in \citere{Dittmaier:2009cr} (\cf Fig.~12 therein).
Let us first analyze the right column of
\reffi{fig:inc-m2lep-highres}.  The weak corrections stay at the $5\%$
level and change sign in the vicinity of the resonance.  This can be
understood from the fact that in the vicinity of the Z~resonance there
are two different types of contributions: corrections to the resonant
part of the squared amplitude, and corrections to the interference of
the resonant and non-resonant parts of the amplitude. The former give
a constant offset of $\sim -5\%$, while the latter are proportional to
$(M_{\mu^+\mu^-}^2-M_\PZ^2)$ and thus change sign at the Z~resonance.
This qualitatively explains the observed sign change of the purely
weak corrections which is slightly shifted below the resonance due to
the negative offset mentioned above. The corrections are to a large
extent equal for the mixed and equal flavour case with minor
deviations of $\sim 1\%$ in the far off-shell region. Including also
the photonic corrections, we observe in both cases the typical
radiative tail due to final-state radiation effects below the Z
resonance, similar to what has been observed in the four-lepton
invariant mass.  In the high-energy spectrum of the steeply falling
invariant-mass distribution, shown in \reffi{fig:inc-m2lep-lowres},
both photonic and purely weak corrections are negative.  The EW
corrections for the invariant mass $M_{\mu^+\mu^-}^{\rm lead}$ of the
leading $\mu^+\mu^-$ pair differ significantly from the mixed-flavour
case due to the large differences at LO. At the peak around $45\GeV$,
the purely weak corrections basically vanish, which is consistent with
the four-lepton invariant-mass distribution in \reffi{fig:inc-m4lep}
where the purely weak corrections vanish at $M_{4\ell}=M_\PZ$. At the
resonance $M_{2\ell}=M_\PZ$, the weak corrections are equal for the
$[2\mu2\Pe]$ and the $[4\mu]$ case because the dominant contribution
where the leading and the subleading lepton pairs are both close to
the resonance is not sensitive to the lepton pairing (note the ratio
of $({\rm d}\sigma[2\mu2\Pe]/{\rm d}M_{2\ell})/(2{\rm
  d}\sigma[4\mu]/{\rm d}M_{2\ell})\approx 0.5$ at the resonance in the
lowest panel and the discussion for the subleading case above the
resonance).  The weak corrections stay always below $5\%$ in absolute
size.  The photonic corrections exhibit an additional radiative tail
below the peak around $45\GeV$. The radiative tail below the
Z~resonance is less pronounced due to the missing resonance
enhancement by the subleading lepton pair.

In \reffi{fig:inc-m2lep-lowres}~(right-hand side) we show the
distribution in the transverse momentum of the (subleading)
$\mu^+\mu^-$ pair, which can be compared with the distribution of the
Z-boson transverse momentum in on-shell calculations.  Since the two
lepton pairs are back to back at LO, the transverse-momentum
distribution depends on the choice of the lepton pair only very
weakly. The interference effect of a few percent is only visible for
small $p_{\rm T,\mu^+\mu^-}$. The EW corrections grow up to $-45\%$
for $p_{\rm T,\mu^+\mu^-}=800\GeV$, while the photonic corrections
stay at the level of $1\%$.  Choosing the $\mu^+\mu^-$ pair from
$[2\mu2\Pe]$, makes it possible to compare the $p_{\rm T,\mu^+\mu^-}$
distribution of our off-shell calculation with the $p_{\rm T,Z}$
distributions for ZZ~production with on-shell Z~bosons discussed in
\citeres{Bierweiler:2013dja,Baglio:2013toa}.  However, as mentioned
above, it should be kept in mind that it cannot be expected to find
perfect agreement because of the differences in the event selection,
which is based on final-state leptons in our calculation, and the
absence of corrections to the Z-boson decays in the on-shell
calculations.  References~\cite{Bierweiler:2013dja,Baglio:2013toa}
state about $-40\%$ at $p_{\rm T,Z}=700 \GeV$ at $\sqrt{s}=14\TeV$,
which agrees with our result for $\sqrt{s}=13\TeV$ at the percent
level in spite of the different setups.

\bfi
\begin{center}
\begin{minipage}{0.49\textwidth}
 \includegraphics[width=\textwidth]{finalplots/{{ATLASincl.13TeV.pt.mu+1.ratio.4mu.2mu2e.collsafe}}}
\end{minipage}
\begin{minipage}{0.49\textwidth}
 \includegraphics[width=\textwidth]{finalplots/{{ATLASincl.13TeV.pt.mu+2.ratio.4mu.2mu2e.collsafe}}}
\end{minipage}
\end{center}
\caption{Transverse-momentum distribution of the $\mu^+$
  (upper panels), corresponding EW corrections (middle panels), and
  ratio of the $[2\mu2\Pe]$ and $[4\mu]$ final states (lower panels)
  in the inclusive setup. The left panels compare the leading $\mu^+$
  from the $[4\mu]$ final state with the $\mu^+$ from the $[2\mu2\Pe]$
  final state, while the panels in the right column show the
  corresponding comparison with the subleading~$\mu^+$.}
\label{fig:inc-ptmup}
\efi
Figure~\ref{fig:inc-ptmup} shows the transverse-momentum distribution
of the $\mu^+$. The left panels compare the leading $\mu^+$ from the
$[4\mu]$ final state with the $\mu^+$ from the $[2\mu2\Pe]$ final
state, the panels in the right column show the corresponding
comparison with the subleading $\mu^+$. Recall that our ordering of
muons into ``leading" and ``subleading" corresponds to the ordering
with respect to the distances of the virtualities of $\mu^+\mu^-$
pairs from the Z~resonance, as described in
\refse{sec:observablesAndCuts}, but not to the muon $p_{\rm T}$, which
is frequently used as well.  We observe again that the observable is
very sensitive to the event selection with characteristic differences
between leading and subleading leptons.  Especially at high transverse
momenta $p_{\rm T}$, the spectrum of the leading muon in $[4\mu]$ is
suppressed with respect to the spectrum of $p_{\rm T}$ in
$[2\mu2\Pe]$, while the spectrum of the subleading muon in $[4\mu]$ is
enhanced.  The difference can be traced back to the impact of the ZZ
signal and background contributions at large transverse momenta: The
leading lepton belongs to the ``more resonant" Z~boson, and therefore,
the contribution is in general dominated by the doubly-resonant signal
contributions [\cf \reffi{fig:born}(a)]. The main effect of the
background contribution in \reffi{fig:born}(b) for large
$p_{\mathrm{T}}$ arises when the $\mu^+$ is back-to-back with the
three other charged leptons. As already observed for the related
process of $\Pp\Pp\to \PW \PW \to$~leptons \cite{Biedermann:2016guo},
the impact of background diagrams on the $p_{\rm T}$ spectrum of a
single lepton can be as large as the doubly-resonant contribution in
the TeV range. Since there is no preselection of the $\mu^+$ in the
$[2\mu2\Pe]$ final state with respect to the resonance, the $p_{{\rm
    T}\mu^+}$ spectrum of $[2\mu2\Pe]$ lies between the spectra of the
leading and subleading muons. This behaviour is also reflected in the
size of the purely weak corrections: Since the Sudakov enhancement is
larger in doubly-resonant contributions than in singly-resonant
contributions, the corrections reach at high $p_{\rm T}$ about $-45\%$
for the leading $\mu^+$, about $-35\%$ for the $\mu^+$ of the mixed
flavour case, and about $-30\%$ in case of the subleading $\mu^+$.
The photonic corrections give an almost constant negative offset of
$-1\%$ to $-2\%$ for the mixed-flavour final state. The shape of the
photonic corrections in the equal-flavour final state is very similar
to the mixed-flavour case. For the subleading lepton, they amount to
$-1.5\%$ to $-3\%$, while for the leading lepton they stay between
$-1\%$ and $-0.5\%$.

\bfi
\begin{center}
\begin{minipage}{0.49\textwidth}
 \includegraphics[width=\textwidth]{finalplots/{{ATLASincl.13TeV.rapidity.mu+1.ratio.4mu.2mu2e.collsafe}}}
\end{minipage}
\begin{minipage}{0.49\textwidth}
 \includegraphics[width=\textwidth]{finalplots/{{ATLASincl.13TeV.rapidity.mu+2.ratio.4mu.2mu2e.collsafe}}}
\end{minipage}
\end{center}
\caption{Rapidity distribution of the $\mu^+$
  (upper panels), corresponding EW corrections (middle panels), and
  ratio of the $[2\mu2\Pe]$ and $[4\mu]$ final states (lower panels) in
  the inclusive setup. The left panels compare the leading $\mu^+$
  from the $[4\mu]$ final state with the $\mu^+$ from the $[2\mu2\Pe]$
  final state, the panels in the right column show the corresponding
  comparison with the subleading~$\mu^+$.}
\label{fig:inc-rapidity}
\efi
\vspace{1em}
\noindent
{\it Rapidity and angular distributions}\\*[.5em]
The rapidity distributions in \reffi{fig:inc-rapidity} do not show any
significant dependence on the lepton pairing except for a small effect
at the percent level in the forward direction with rapidities
$|y_{\mu^+}|>2$.  The EW corrections are independent of the lepton
pairing as well, and their size is almost equal for both final states.
The purely photonic corrections give, in good approximation, a
constant negative offset of roughly $-1\%$, reflecting the results of
the integrated cross section in \refta{tab:inc-xsec}. The impact of
the weak corrections is numerically largest in the central region with
about $-4.7\%$ and less negative in the forward direction with about
$-3\%$.

\bfi
\begin{center}
\begin{minipage}{0.49\textwidth}
 \includegraphics[width=\textwidth]{finalplots/{{ATLASincl.13TeV.azimuthdis.mu+mu-11.ratio.4mu.2mu2e.collsafe}}}
\end{minipage}
\begin{minipage}{0.49\textwidth}
 \includegraphics[width=\textwidth]{finalplots/{{ATLASincl.13TeV.azimuthdis.mu+mu-22.ratio.4mu.2mu2e.collsafe}}}
\end{minipage}
\end{center}
\caption{Azimuthal-angle difference  between muons within the
  $\mu^+\mu^-$ pair (upper panels), corresponding EW corrections
  (middle panels), and ratio of the $[2\mu2\Pe]$ and $[4\mu]$ final
  states (lower panels) in the inclusive setup.  The left panels
  compare the leading $\mu^+\mu^-$~pair from the $[4\mu]$ final state
  with the $\mu^+\mu^-$~pair of the $[2\mu2\Pe]$ final state, the
  panels in the right column show the corresponding comparison with
  the subleading lepton pair.}
\label{fig:inc-azimuthaldis}
\efi
In \reffi{fig:inc-azimuthaldis}, the distribution in the
azimuthal-angle distance between the muons in the $\mu^+\mu^-$~pair is
shown. We observe a maximum for $\Delta\phi_{\mu^+\mu^-}\to \pi$ that
is reached in good approximation independently of the final state and
the lepton pairing. This can be explained as follows: The
azimuthal-angle-distance distribution is dominated in the whole range
by events in the low-energy region around the threshold at
$M_{4\ell}=2M_\PZ$ where the cross section receives the largest
contribution from doubly-resonant contributions.  Moreover, the
$t$-channel nature of the doubly-resonant diagrams favours small
transverse momenta of the Z~bosons. As can also be seen in
\reffi{fig:inc-ptmup}, most of the leptons have $p_{\rm T}\lesssim
M_\PZ/2$ as a result of the decay of the Z~bosons that move slowly in
the transverse plane, \ie the Z~bosons decay almost isotropically in
the transverse plane, without a large influence of boost effects from
a transverse momentum of the Z~boson.  For small
$\Delta\phi_{\mu^+\mu^-}$, the behaviour of the distribution is
completely different. The $\mu^+\mu^-$ pair in $[2\mu2\Pe]$ as well as
the subleading $\mu^+\mu^-$ pair in $[4\mu]$ show some enhancement
near $\Delta\phi_{\mu^+\mu^-}\sim 0.2$, while this enhancement is
absent for the leading $\mu^+\mu^-$ pair. This is a result of events
with small $\mu^+\mu^-$ invariant masses $M_{\mu^+\mu^-}< M_\PZ$.
Owing to the $\mu^+\mu^-$ pairing with respect to their invariant
mass, low-mass $\mu^+\mu^-$ pairs rarely occur as leading pairs, as
can be seen in \reffi{fig:inc-m2lep-highres}. On the other hand,
low-mass $\mu^+\mu^-$ pairs receive a much larger contribution from
virtual photon $(\gamma^\star)$ exchange than nearly resonant
$\mu^+\mu^-$ pairs, leading to the observed peak structures at
$\Delta\phi_{\mu^+\mu^-}\sim 0.2$. Note that these peaks are truncated
on their left side by the lepton separation cut of
Eq.~\refeq{eq:inc-cuts2}.

In the dominant region of large $\Delta\phi_{\mu^+\mu^-}$, the weak
corrections resemble the size observed for the integrated cross
sections above. For smaller $\Delta\phi_{\mu^+\mu^-}$ the negative
corrections tend to increase in size, because this region in
$\Delta\phi_{\mu^+\mu^-}$ on average requires more scattering energy
to yield boost effects that turn the $\mu^+$ and $\mu^-$ directions
away from the back-to-back configuration preferred at low energies.
The effect of increasing negative weak corrections for smaller
$\Delta\phi_{\mu^+\mu^-}$ is to some extent balanced in the
$[2\mu2\Pe]$ case and for the subleading $\mu^+\mu^-$ pair in
$[4\mu]$, because the weak corrections to the significantly
contributing $\PZ\gamma^\star$ production diagrams are smaller than
the ones for ZZ production (\cf results on $\PZ\gamma$ production
shown in \citere{Denner:2015fca}).  The photonic corrections are
generically small for the $\Delta\phi_{\mu^+\mu^-}$ distribution owing
to the absence of collinear enhancements, because collinear
final-state radiation does not change the directions of the radiating
leptons significantly. The photonic corrections are most sizeable at
$\Delta\phi_{\mu^+\mu^-}\to \pi$ with $-1.2\%$,$-1\%$, and $ -1.5\%$
for the mixed-flavour case, the leading lepton pair, and the
subleading lepton pair, respectively.  They decrease towards
$\Delta\phi_{\mu^+\mu^-}\to 0$ where they reach $-0.3\%$, $-0.7\%$,
and $+0.2\%$, respectively.
\bfi
\begin{center}
\begin{minipage}{0.49\textwidth}
 \includegraphics[width=\textwidth]{finalplots/{{ATLASincl.13TeV.ZZdecayangle_ZZangle.ratio.4mu.2mu2e.collsafe}}}
\end{minipage}
\end{center}
\caption{Angle between the two Z-boson decay planes in the CM system for the
  unequal-flavour $[2\mu2\Pe]$ and for the leading and subleading
  $\mu^+\mu^-$ pair of the equal-flavour $[4\mu]$ final state (upper
  panels), corresponding EW corrections 
  (middle panels), and ratio of the $[2\mu2\Pe]$ and $[4\mu]$ final
  states (lower panels) in the inclusive setup.}
\label{fig:inc-ZbosonDecayPlaneAngle}
\efi

Figure \ref{fig:inc-ZbosonDecayPlaneAngle} shows the distribution in
the angle $\phi$ between the two Z-boson decay planes for the
unequal-flavour $[2\mu2\Pe]$ final state and the corresponding angle
defined by the leading and subleading $\mu^+\mu^-$ pairs of the
equal-flavour $[4\mu]$ final state.
The angle $\phi$ is defined 
in the CM system of the four final-state leptons
by
\begin{eqnarray}
\cos{\phi} &=& 
\frac{({\bf k}_{\ell_1^+\ell_1^-}\times{\bf k}_{\ell_1^+})({\bf k}_{\ell_1^+\ell_1^-}\times{\bf k}_{\ell_2^+})}
     {|{\bf k}_{\ell_1^+\ell_1^-}\times{\bf k}_{\ell_1^+}||{\bf k}_{\ell_1^+\ell_1^-}\times{\bf k}_{\ell_2^+}|},
\nonumber\\
\sgn(\sin{\phi}) &=& 
\sgn\{{\bf k}_{\ell_1^+\ell_1^-}\cdot[({\bf k}_{\ell_1^+\ell_1^-}\times{\bf k}_{\ell_1^+})\times
                        ({\bf k}_{\ell_1^+\ell_1^-}\times{\bf k}_{\ell_2^+})]\},
\label{eq:phipr}
\end{eqnarray}  
where ${\bf k}_{\ell_1^+\ell_1^-}={\bf k}_{\ell_1^+}+{\bf
  k}_{\ell_1^-}$, and ${\bf k}_{i}$ denote the spatial parts of the
four lepton momenta $k_i$,
$i=\{{\ell_1^+},{\ell_1^-},{\ell_2^+},{\ell_2^-}\}$ of the $\mmee$ and
$\mmmm$ final states in the CM system.  The dips in the
distribution for coinciding decay planes, \ie for $\phi\sim0$ and
$\phi\sim\pi$, are a consequence of the lepton-separation cuts in
Eq.~\refeq{eq:inc-cuts2}. These cuts remove collinear lepton
configurations where the decay planes tend to be coplanar.  The local
minima around $\phi=\pi/2$ and $\phi=3\pi/2$ can be understood from
the superposition of contributions with different lepton helicities:
If the two equally charged final-state leptons do have the same
helicity, the distribution shows a maximum at $\phi=0$ and a minimum
at $\phi=\pi$, and vice versa for opposite lepton helicities. The
corresponding distribution in $\PW^+\PW^-$ pair production (c.f.\ 
Fig.~16 in \citere{Denner:2005es}) exhibits a maximum at $\phi=0$ and
a minimum at $\phi=\pi$, as expected for purely left-handed leptons.
We observe an enhancement of the ratio $[2\mu2\Pe]/(2[4\mu])$ above
one for small angles $\phi$ and a reduction for angles $\phi$ close to
$\pi$, both at the level of a few percent. Since an exchange of the
leading and subleading $\mu^+\mu^-$ pair does not affect $\phi$, we
attribute this effect to the interference between the two different
classes of diagrams in $[4\mu]$ with different $\PZ\to\mu^+\mu^-$
pairings.  For a qualitative understanding of the interference
pattern, it is instructive to consider the limit of nearly coplanar
Z~decays $(\phi\to 0,\pi)$.  Since photon emission effects are not
dominating this observable, the majority of the events have ${\bf
  k}_{\mu_1^+\mu_1^-}\approx -{\bf k}_{\mu_2^+\mu_2^-}$, whose
direction defines the intersection line of the two decay planes.  In
the vicinity of the coplanar configurations this line divides the
event plane into two half planes.  According to the definition
\refeq{eq:phipr} of $\phi$, the two vectors ${\bf
  k}_{\mu_1^+\mu_1^-}\times{\bf k}_{\mu_1^+}$ and ${\bf
  k}_{\mu_1^+\mu_1^-}\times{\bf k}_{\mu_2^+}$ are parallel for
$\phi\to0$ and antiparallel for $\phi\to\pi$, i.e.\ ${\bf
  k}_{\mu_1^+}$ and ${\bf k}_{\mu_2^+}$ lie in the same half plane for
$\phi\to0$, but in different half planes for $\phi\to\pi$.  On
average, we thus find more parallel ${\mu_1^+\mu_2^+}$ pairs for
$\phi\to0$ and more antiparallel pairs for $\phi\to\pi$.  Since the
matrix elements are antisymmetrized with respect to the exchange of
$\mu^+_1\leftrightarrow\mu^+_2$ or $\mu^-_1\leftrightarrow\mu^-_2$, a
destructive interference is favoured for $\phi\to0$ in $[4\mu]$,
leading to the observed enhancement in the $[ 2\mu2\Pe]/(2[4\mu])$
ratio.  
This effect is not changed by the EW corrections. The weak
corrections distort the $\phi$ distribution by about $3\%$, while the
photonic corrections only uniformly contribute by $-1\%$.

\vspace{1em}
\noindent
{\it Collinear-safe versus collinear-unsafe observables}
\\[.5em]
\bfi
\begin{center}
\begin{minipage}{0.49\textwidth}
 \includegraphics[width=\textwidth]{finalplots/{{ATLASincl.13TeV.invmass.4lep.4mu.2mu2e.highres.safe-unsafe}}}
\end{minipage}
\begin{minipage}{0.49\textwidth}
 \includegraphics[width=\textwidth]{finalplots/{{ATLASincl.13TeV.invmass.mu+mu-.2mu2e.lowres.safe-unsafe}}}
\end{minipage}
\end{center}
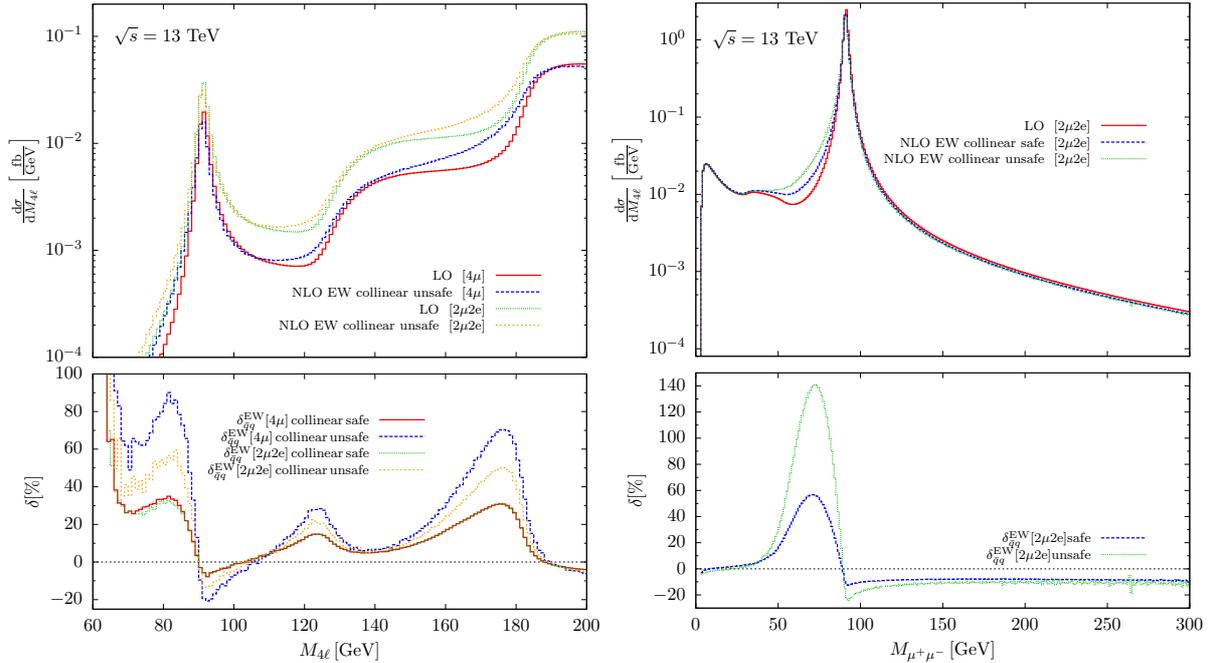
\caption{Comparison of different photon recombination schemes for the
  four-lepton and two-lepton invariant-mass distributions. The upper
  panels show the absolute distributions and the lower panels the
  relative EW corrections.}
\label{fig:inc-save-unsafe}
\efi
In \reffi{fig:inc-save-unsafe}, the different recombination schemes
for the muon are illustrated for the four-lepton and two-lepton
invariant masses.  The recombination procedure only affects the
photonic corrections.  As a general pattern, all the radiative tails
induced by final-state radiation off the charged leptons are strongly
enhanced if collinear photons are not recombined with muons.  The
enhancement is due to the fact that the collinear logarithms are
regularized by the muon mass rather than the size of the recombination
cone. The effects can be best isolated in the $M_{4\ell}$
invariant-mass distribution (left panels of
\reffi{fig:inc-save-unsafe}) which is not sensitive to the lepton
pairing. While the absolute prediction is only shown for the
collinear-unsafe case for the mixed- and equal-flavour final states,
the relative EW corrections are plotted both for the collinear-safe
and -unsafe cases. The results illustrate the impact of the number of
muons excluded from recombination to the distribution: The maximum of
the radiative tail below the ZZ threshold increases from about $+30\%$
with full recombination to more than $+50\%$ for excluding one muon
pair ($[2\mu2\Pe]$) up to about $+70\%$ by excluding both muon pairs
($[4\mu]$). For $[4\mu]$, the increase is twice as large as for
$[2\mu2\Pe]$, since the recombination effect scales with the number of
collinear cones that are subject to the changes in the recombination.
A similar behaviour is found for the other radiative tails at smaller
values of $M_{4\ell}$. An even stronger enhancement can be seen at the
invariant mass of the $\mu^+\mu^-$ pair (right panels) below the
Z~resonance where the relative correction increases from almost
$+60\%$ to $+140\%$. Note that above the resonance the effect from the
collinear-unsafe treatment pushes the negative collinear-safe
\looseness -1 corrections even more negative.

\subsection{Results on differential cross sections in the Higgs-specific setup}

{\it Invariant-mass and transverse-momentum distributions} 
\\[.5em]
The production of Z-boson pairs at the LHC is interesting not only per
se, as a signal process, but also constitutes 
an important irreducible background to Higgs production in the $\PH\to
\PZ\PZ^\star$ decay mode. In order to assess the impact of this
background on Higgs analyses, we impose the Higgs-specific cuts of
Eqs.~(\ref{eq:hcuts1})--(\ref{eq:hcuts3}) in addition to the inclusive
cut of Eq.~(\ref{eq:inc-cuts2}).  In \citere{Biedermann:2016yvs} we
already presented some important results of this study, however,
restricted to the unequal-flavour final-state $[2\mu2\Pe]$ and
ignoring photon-induced channels.  In the following we continue the
discussion started there by comparing results for the $[2\mu2\Pe]$ and
$[4\mu]$ final states and considering further observables.

Figure~\ref{fig:higgs-m4lep} 
\bfi
\begin{center}
\begin{minipage}{0.49\textwidth}
 \includegraphics[width=\textwidth]{finalplots/{{HiggsBG.13TeV.invmass.4lep.ratio.4mu.2mu2e.highres.collsafe}}}
\end{minipage}
\begin{minipage}{0.49\textwidth}
  \includegraphics[width=\textwidth]{finalplots/{{HiggsBG.13TeV.invmass.4lep.ratio.4mu.2mu2e.lowres.collsafe}}}
\end{minipage}
\end{center}
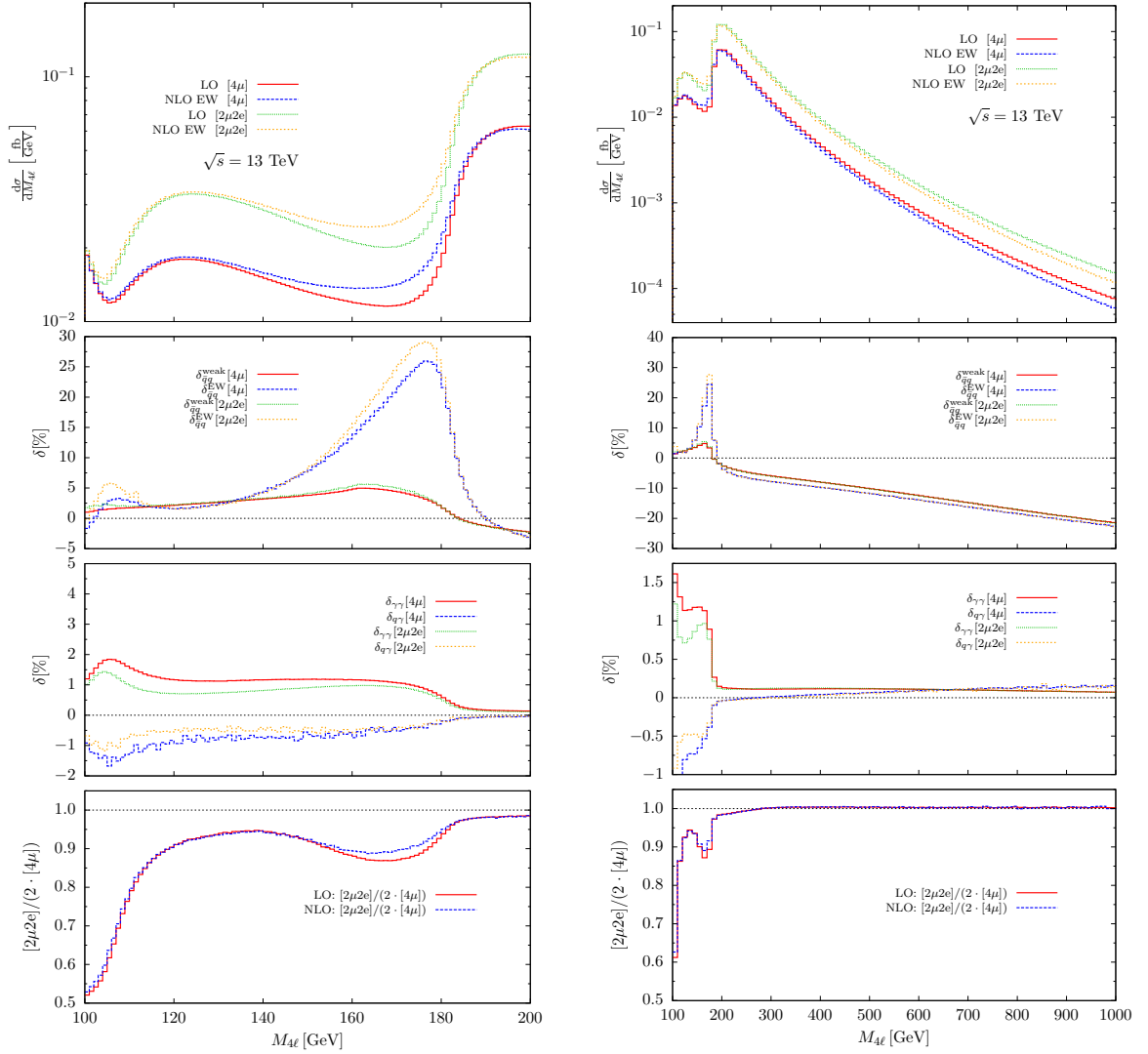
\caption{Invariant-mass distribution of the four-lepton system (upper panels),
  corresponding EW corrections (2nd panels from above), photonic
  contributions (third panels from above) for the unequal-flavour
  $[2\mu2\Pe]$ and the equal-flavour $[4\mu]$ final states in the
  Higgs-specific setup. The lower panels show the ratio of the $[2\mu2\Pe]$
  and $[4\mu]$ final states.} 
\label{fig:higgs-m4lep}
\efi
illustrates the invariant-mass distribution of the four-lepton system
at LO and the corresponding NLO EW corrections for both the
$[2\mu2\Pe]$ and the $[4\mu]$ final states.  In each case, we observe
a steep shoulder at the Z-boson pair production threshold at about
$M_{4\ell}=2M_\PZ\approx182\GeV$, which gives rise to a large
radiative tail in the photonic corrections at smaller invariant
masses.  Though smaller in magnitude, a similar effect can be observed
at around $M_\PZ+2p_{\rm T,{\rm min}}\approx103\GeV$ which is due to
the transverse-momentum and invariant-mass cuts we impose on the
charged leptons.  Like in the inclusive setup, both the purely weak
and the photonic corrections exhibit a sign change at the pair
production threshold around $M_{4\ell}=2M_\PZ$. The pattern of the EW
corrections above the ZZ threshold is very similar to the inclusive
setup with at most permille level differences between the $[4\mu]$ and
the $[2\mu2\Pe]$ case. The photonic corrections decrease in absolute
size from approximately $-2\%$ at the threshold to about $-1\%$ at
$1\TeV$.  The purely weak corrections constantly increase in absolute
size reaching about $-20\%$ at $1\TeV$. Also in the
off-shell-sensitive region below the pair production threshold, the
difference between the $[4\mu]$ and the $[2\mu2\Pe]$ cases in the
purely weak corrections is below the percent level. The radiative
tails in the photonic corrections are up to $5\%$ larger in the
mixed-flavour case.  In contrast to the inclusive setup, the
phase-space cuts of the Higgs-specific setup introduce a dependence on
the lepton pairing even in otherwise symmetric observables like the
four-lepton invariant mass. The difference seen in the photonic
corrections is thus due to both the lepton pairing and the
interference effects. At the Higgs-boson mass $M_{4\ell}=M_\PH$, the
differences of the EW corrections with respect to the final states
are, however, entirely negligible.  The significant differences
between the $[4\mu]$ and the $[2\mu2\Pe]$ case in the
off-shell-sensitive region are, like in the inclusive case, a priori a
LO effect. Note that the non-trivial sign change of the photonic
corrections leads to significant cancellations between opposite-sign
contributions below and above the ZZ threshold resulting in
sub-permille effects in the total cross section (\cf
\refta{tab:inc-xsec}), although the individual photonic corrections
can be sizable in distributions.

We also show the photon-induced contribution to the four-lepton
invariant-mass distribution in the third panels from above in
\reffi{fig:higgs-m4lep}. Above the ZZ production threshold the
corrections are at the level of one permille. In the off-shell region,
the $q\gamma$ and the $\gamma\gamma$~contribution are opposite in
sign, at the level of $1\%$, and compensate each other to a large
extent. The overall impact remains at the sub-percent level. We do not
show the photon-induced corrections separately in the following plots.

In \reffi{fig:higgs-mmm-highres} the invariant-mass distribution of
the $\mm$ system is shown for the $[2\mu2\Pe]$ final state, as well as
the ones for the leading and the subleading $\mm$ systems of the
$[4\mu]$ final state.
\bfi
\begin{center}
\begin{minipage}{0.49\textwidth}
 \includegraphics[width=\textwidth]{finalplots/{{HiggsBG.13TeV.invmass.mu+mu-11.ratio.4mu.2mu2e.highres.collsafe}}}
\end{minipage}
\begin{minipage}{0.49\textwidth}
 \includegraphics[width=\textwidth]{finalplots/{{HiggsBG.13TeV.invmass.mu+mu-22.ratio.4mu.2mu2e.highres.collsafe}}}
\end{minipage}
\caption{Invariant $\mu^+\mu^-$-mass distribution (upper panels),
  corresponding EW corrections (middle panels), and ratio of the
  $[2\mu2\Pe]$ and $[4\mu]$ final states (lower panels) in the
  Higgs-specific  setup.  In the left column the equal-flavour case is
  binned with respect to the leading lepton pair, while the right
  column shows results for the subleading one.\label{fig:higgs-mmm-highres}
}
\end{center}  
\efi 
Due to the cuts of Eq.~(\ref{eq:hcuts1}), the invariant mass of the
leading muon pair in the equal-flavour final state is restricted to
the range of $40{-}120\GeV$. This cut leads in the $[2\mu2\Pe]$ final
state to a little bump at $40\GeV$.  Moreover, the local maximum near
$\MZ/2$ in the $M^{\mathrm{lead}}_{\mm}$ distribution of the inclusive
setup is absent for the Higgs-specific setup, because the
invariant-mass cut $M_{4\ell}>100\GeV$ in Eq.~(\ref{eq:hcuts2})
entirely removes the $s$-channel resonance at $M_{4\ell}=M_\PZ$.  Near
the Z~resonance the photonic and weak corrections are very similar to
the results in the inclusive setup (\cf
\reffi{fig:inc-m2lep-highres}).  The distribution peaks at the
resonance, $M_{\mm}\sim \MZ$, and receives large photonic corrections
below that are due to final-state radiation effects.  The weak
corrections, on the other hand, are of the order of $5\%$ and give
rise to a change in sign near the Z-boson resonance.  Above the
resonance the EW corrections are qualitatively similar to the ones in
the inclusive setup for both the leading and subleading $\mu^+\mu^-$
pair.  Below $M_{\mm}\approx 60\GeV$, on the other hand, the missing
$s$-channel resonance at $M_{4\ell}=M_\PZ$ leads to significant
changes.  The difference is, as expected, most prominent in the
leading lepton pair where the local minimum of the weak corrections at
$45 \GeV$ and the entire additional radiative tail of the photonic
corrections are removed.  While the EW corrections show sizeable
deviations between the mixed- and equal-flavour final states, the main
differences are LO effects that can be attributed to the cuts and the
lepton pairing.

Figure~\ref{fig:higgs-ptmup} 
%%%
%
\bfi
\begin{center}
\begin{minipage}{0.49\textwidth}
 \includegraphics[width=\textwidth]{finalplots/{{HiggsBG.13TeV.pt.mu+1.ratio.4mu.2mu2e.collsafe}}}
\end{minipage}
\begin{minipage}{0.49\textwidth}
 \includegraphics[width=\textwidth]{finalplots/{{HiggsBG.13TeV.pt.mu+2.ratio.4mu.2mu2e.collsafe}}}
\end{minipage}
\end{center}
\caption{Transverse-momentum distribution of the $\mu^+$
(upper panels), corresponding EW corrections (middle panels), and
ratio of the $[2\mu2\Pe]$ and $[4\mu]$ final states (lower panels) in the
Higgs-specific setup. The left panels compare the leading $\mu^+$ from the
$[4\mu]$ final state with the $\mu^+$ from the $[2\mu2\Pe]$ final
state, while the panels in the right column show the corresponding
comparison with the subleading~$\mu^+$.}
\label{fig:higgs-ptmup}
\efi
depicts the transverse-momentum distribution of the $\mu^+$ in the
$[2\mu2\Pe]$ final state together with the leading and the subleading
$\mu^+$ of the $[4\mu]$ final state, respectively. We once again 
remind the reader that the classification of leptons as
``leading'' or ``subleading'' refers to the criteria of
Eq.~(\ref{eq:hcuts1}), \ie the leading muon is not necessarily the
muon of highest transverse momentum, but stems from the $\mm$~pair
with the invariant mass closest to the mass of the Z~boson. We find that,
in contrast to the inclusive setup illustrated in
\reffi{fig:inc-ptmup}, the weak corrections to the transverse momenta
are very similar in size and shape for the equal- and the
unequal-flavour cases. They become large and negative in the tails,
amounting to $-50\%$ already at about $800\GeV$. 
This is mainly a result of the suppression of background diagrams of
the type shown in \reffi{fig:born}(b), which can already be seen from
the suppression of the absolute LO cross section at large transverse
momenta (\cf \reffi{fig:inc-ptmup} and related discussion there).   
The impact of photonic corrections is at the level of one percent for
small transverse momenta and even smaller for large ones for both
leptonic final states.

\vspace{1em}
\noindent
{\it Rapidity and angular distributions}
\\*[.5em]
The rapidity distributions of the $\mu^+$ and the corresponding EW
corrections, shown in \reffi{fig:higgs-rapidity}, do not change very
much when going from the inclusive to the Higgs-specific setup.
\bfi
\begin{center}
\begin{minipage}{0.49\textwidth}
 \includegraphics[width=\textwidth]{finalplots/{{HiggsBG.13TeV.rapidity.mu+1.ratio.4mu.2mu2e.collsafe}}}
\end{minipage}
\begin{minipage}{0.49\textwidth}
 \includegraphics[width=\textwidth]{finalplots/{{HiggsBG.13TeV.rapidity.mu+2.ratio.4mu.2mu2e.collsafe}}}
\end{minipage}
\end{center}
\caption{Rapidity distribution of the $\mu^+$
  (upper panels), corresponding EW corrections (middle panels), and
  ratio of the $[2\mu2\Pe]$ and $[4\mu]$ final states (lower panels) in
  the Higgs-specific setup. The left panels compare the leading $\mu^+$
  from the $[4\mu]$ final state with the $\mu^+$ from the $[2\mu2\Pe]$
  final state, the panels in the right column show the corresponding
  comparison with the subleading~$\mu^+$.}
\label{fig:higgs-rapidity}
\efi
The only visible changes are the constant offsets in the relative
corrections that can already be observed for the integrated cross
sections given in \refta{tab:inc-xsec}.

Figure~\ref{fig:higgs-azimuthaldis} illustrates the distribution in
the azimuthal-angle difference of the leading and the subleading
$\mm$~pair in the $[4\mu]$ final state together with the respective
distribution for the $\mm$~pair of the $[2\mu2\Pe]$ final state.
\bfi
\begin{center}
\begin{minipage}{0.49\textwidth}
 \includegraphics[width=\textwidth]{finalplots/{{HiggsBG.13TeV.azimuthdis.mu+mu-11.ratio.4mu.2mu2e.collsafe}}}
\end{minipage}
\begin{minipage}{0.49\textwidth}
 \includegraphics[width=\textwidth]{finalplots/{{HiggsBG.13TeV.azimuthdis.mu+mu-22.ratio.4mu.2mu2e.collsafe}}}
\end{minipage}
\end{center}
\caption{Azimuthal-angle difference  between muons within the
  $\mu^+\mu^-$ pair (upper panels), corresponding EW corrections
  (middle panels), and ratio of the $[2\mu2\Pe]$ and $[4\mu]$ final
  states (lower panels) in the Higgs-specific setup.  The left panels
  compare the leading $\mu^+\mu^-$~pair from the $[4\mu]$ final state
  with the $\mu^+\mu^-$~pair of the $[2\mu2\Pe]$ final state, the
  panels in the right column show the corresponding comparison with
  the subleading lepton pair.}
\label{fig:higgs-azimuthaldis}
\efi
Interestingly, the peak structure observed for the analogous
distributions in the inclusive setup shown in
\reffi{fig:inc-azimuthaldis} is absent in the Higgs-specific setup.
As noted above, the enhancements in the inclusive setup at
$\Delta\phi_{\mm}\sim 0.2$ are mostly due to $\mm$ pairs of low
invariant mass enhanced by the photon pole. The Higgs-specific
selection cuts applied in the current setup remove such contributions,
leaving us with azimuthal-angle distributions that still exhibit a
similar rise towards $\Delta\phi_{\mm}\to \pi$, but no longer peak at
low values of $\Delta\phi_{\mm}$.  Apart from the peak structure, the
impact of weak corrections on normalization and shape of the
azimuthal-angle differences in the Higgs setup is similar in size as
in the inclusive setup.  Purely photonic corrections are even more
suppressed than in the inclusive case.  In the Higgs-specific scenario
the fraction of events with leading muon pairs close to the
Z~resonance is enhanced, while the one for subleading muon pairs is
reduced (compare \reffis{fig:inc-m2lep-highres} and
\ref{fig:higgs-mmm-highres}).  As a consequence, the distribution of
the leading muon pair is enhanced compared to the unequal-flavour case
for large $\Delta\phi_{\mm}$, while the one of the subleading lepton
pair is reduced. For small azimuthal-angle differences the situation
is reversed.

\bfi
\begin{center}
\begin{minipage}{0.49\textwidth}
 \includegraphics[width=\textwidth]{finalplots/{{HiggsBG.13TeV.ZZdecayangle_ZZangle.ratio.4mu.2mu2e.collsafe}}}
\end{minipage}
\end{center}
\caption{Angle between the two Z-boson decay planes in the CM system for the
  unequal-flavour $[2\mu2\Pe]$ and for the leading and subleading
  $\mu^+\mu^-$ pair of the equal-flavour $[4\mu]$ final state (upper
  panels), corresponding EW corrections 
  (middle panels), and ratio of the $[2\mu2\Pe]$ and $[4\mu]$ final
  states (lower panels) in the Higgs-search setup.}
\label{fig:inc-ZbosonDecayPlaneAngleHiggsBG}
\efi
We show in \reffi{fig:inc-ZbosonDecayPlaneAngleHiggsBG} the
distribution in the angle between the two Z-boson decay planes in the
four-lepton CM system in the Higgs-specific setup.\footnote{The distribution in
  the angle between the two \PZ-boson decay planes shown in Fig.~3 of
  \citere{Biedermann:2016yvs} is not directly comparable to the
  distribution shown in \reffi{fig:inc-ZbosonDecayPlaneAngleHiggsBG}. In
  \citere{Biedermann:2016yvs} the angle $\phi$ has been calculated
  from the lepton momenta in the laboratory system.}
The distribution as well as the EW corrections closely resemble those
of the inclusive setup shown in \reffi{fig:inc-ZbosonDecayPlaneAngle}.
The ratio $[2\mu2\Pe]/(2[4\mu])$ is qualitatively similar, showing
some excess over one for $\phi\to0,2\pi$ and some deficit around
$\phi\sim\pi$, where the latter is, however, more pronounced than in
the inclusive setup.  Owing to the asymmetric treatment of the leading
and subleading $\mu^+\mu^-$ pairs in the Higgs setup, we cannot
attribute the deviations of the ratio from one to interference effects
only.  Finally, we remark that the $\phi$ distribution in the
Higgs-signal process $\PH\to\PZ\PZ^*\to 4\,$leptons looks
qualitatively similar to the distribution of direct $\PZ\PZ$
production shown in \reffi{fig:inc-ZbosonDecayPlaneAngleHiggsBG}, but
the distortions by EW corrections are quite
different~\cite{Bredenstein:2006rh,Bredenstein:2007ec}.

\clearpage

\section{Conclusions}
\label{se:conclusion}

The production of four charged leptons in hadronic collisions at the
LHC is an important process class both for the investigation of the
interactions between the neutral Standard Model gauge bosons and as
background process to searches for new physics and to precision
studies of the Higgs boson.  In the confrontation of experimental data
with theory predictions precision plays a key role. In this paper we
have further improved the theory prediction by calculating the
next-to-leading-order electroweak corrections to the production of
$\mmee$ and $\mmmm$ final states without any kinematical restrictions
on the intermediate states.  Our results are thus accurate to
next-to-leading order in all phase-space regions, no matter whether
they are dominated by two, one, or zero resonant Z~bosons.  Our
numerical discussion of the corrections focuses on two different
event-selection scenarios, one based on typical lepton-identification
criteria only and another one that is specifically designed for
Higgs-boson analyses.  Since the Higgs-boson mass of about $125\GeV$
lies below the Z-pair threshold, the flexibility of our calculation,
allowing intermediate Z~bosons to be far off shell, is essential for
the study of four-lepton production as background to the Higgs-boson
decay $\PH\to\PZ\PZ^\star$.

Extending our earlier study~\cite{Biedermann:2016yvs} of the process
$\ppmmee$, we have investigated further observables and channels with
photons in the initial state and included the process $\ppmmmm$.
Generically, the next-to-leading order electroweak corrections consist
of photonic and purely weak contributions displaying rather different
features. Photonic corrections can grow very large, to several tens of
percent, in particular in distributions where resonances and kinematic
shoulders lead to radiative tails.  These effects are significantly
enhanced when observables within a collinear-unsafe setup are
considered.  While photonic corrections might be well approximated
with QED parton showers, this is not the case for the weak
corrections, which are typically of the size of $-5\%$ at intermediate
energies and grow to multiples of $-10\%$ in the high-energy tails of
invariant-mass and transverse-momentum distributions.  Moreover, the
weak corrections below the ZZ~threshold distort distributions that are
important in Higgs-boson analyses.  On the other hand, contributions
induced by incoming photons, \ie photon--photon and quark--photon
channels, turn out to be phenomenologically unimportant.  Comparing
the results on $\mmee$ and $\mmmm$ final states, we find significant
differences mainly in distributions that are sensitive to the
assignment of $\mu^+\mu^-$ pairs in the $\mmmm$ final state to
intermediate Z~bosons.  Interferences in equal-flavour-lepton final
states lead to deviations of up to $10\%$ from the mixed-flavour case
in off-shell-sensitive phase-space regions. Their effect is, however,
in general hidden in the effects of the selection criteria for the
lepton pairing.  The relative electroweak corrections are widely
insensitive to details of the lepton pairing, \ie the selection of
$\mu^+\mu^-$ pairs affects observables at leading and next-to-leading
order roughly in the same way.

The full calculation is available in the form of a Monte Carlo program
allowing for the evaluation of arbitrary differential cross sections.
The best possible predictions for ZZ~production processes can be
achieved by combining the electroweak corrections of our calculation
with the most accurate QCD predictions available to date. Practically,
this could be achieved, e.g., by reweighting differential
distributions including QCD corrections with electroweak correction
factors.  In this way, an overall accuracy at the percent level can be
achieved for integrated cross sections that are dominated by energy
scales up to a few $100\GeV$, where the theoretical uncertainty is
completely dominated by QCD.  We estimate the contribution of missing
higher-order electroweak corrections on the integrated cross section
to $0.5\%$.  The impact of missing higher-order electroweak
corrections grows in the high-energy tails of transverse-momentum and
invariant-mass distributions where weak Sudakov (and subleading
high-energy) logarithms are known to be large.  In this kinematic
domain, the size of this uncertainty may be estimated by the square of
the relative electroweak correction.  The inclusion of the known
leading two-loop effects or a resummation of logarithmically enhanced
contributions could reduce these theoretical uncertainties. At the
same time, multi-photon emission effects could be systematically taken
into account by structure functions or parton showers.  Such
improvements are, however, left to future studies.  For upcoming
analyses of LHC data, next-to-leading order precision in electroweak
corrections is certainly sufficient, and the remaining electroweak
uncertainties are negligible compared to the larger uncertainties from
missing QCD corrections and from parton distribution functions.

\subsection*{Acknowledgements}

We would like to thank Jochen Meyer for helpful discussions.  The work
of B.B. and A.D. was supported by the German Federal Ministry for
Education and Research (BMBF) under contract no.~05H15WWCA1 and by the
German Science Foundation (DFG) under reference number DE 623/2-1.
S.D.\ gratefully acknowledges support from the DFG research training
group RTG~2044.  The work of L.H. was supported by the grants
FPA2013-46570-C2-1-P and 2014-SGR-104, and partially by the Spanish
MINECO under the project MDM-2014-0369 of ICCUB (Unidad de Excelencia
“Mar\'ia de Maeztu”).  
The work of B.J.\ was supported in part by the Institutional Strategy
of the University of T\"ubingen (DFG, ZUK~63) and in part by the BMBF 
under contract number 05H2015. 
The authors acknowledge support by the state of Baden-W\"urttemberg
through bwHPC and the German Research Foundation (DFG) through grant
no. INST 39/963-1 FUGG.

\clearpage

\bibliographystyle{JHEPmod}             %  JHEP (maximal 4 authors)
\bibliography{bibliography}

\providecommand{\href}[2]{#2}\begingroup\raggedright\begin{thebibliography}{10}

\bibitem{Aad:2014wra}
{\bf ATLAS} Collaboration, G.~Aad et~al., {\it {Measurements of four-lepton
  production at the Z~resonance in pp collisions at $\sqrt{s}=7$ and $8\,$TeV
  with ATLAS}},  {\em Phys. Rev. Lett.} {\bf 112} (2014), no.~23 231806,
  [\href{http://arxiv.org/abs/1403.5657}{{\tt arXiv:1403.5657}}].

\bibitem{Chatrchyan:2012sga}
{\bf CMS} Collaboration, S.~Chatrchyan et~al., {\it {Measurement of the $ZZ$
  production cross section and search for anomalous couplings in $2l2l'$ final
  states in $pp$ collisions at $\sqrt{s}=7\,$TeV}},  {\em JHEP} {\bf 01} (2013)
  063, [\href{http://arxiv.org/abs/1211.4890}{{\tt arXiv:1211.4890}}].

\bibitem{Chatrchyan:2013oev}
{\bf CMS} Collaboration, S.~Chatrchyan et~al., {\it {Measurement of $W^+W^-$
  and ZZ production cross sections in pp collisions at $\sqrt{s} = 8\,$TeV}},
  {\em Phys. Lett.} {\bf B721} (2013) 190--211,
  [\href{http://arxiv.org/abs/1301.4698}{{\tt arXiv:1301.4698}}].

\bibitem{Aaboud:2016urj}
{\bf ATLAS} Collaboration, M.~Aaboud et~al., {\it {Measurement of the $ZZ$
  production cross section in $pp$ collisions at $\sqrt{s} = 8\,$TeV using the
  $ZZ\to\ell^{-}\ell^{+}\ell^{\prime -}\ell^{\prime +}$ and
  $ZZ\to\ell^{-}\ell^{+}\nu\bar{\nu}$ channels with the ATLAS detector}},
  \href{http://arxiv.org/abs/1610.07585}{{\tt arXiv:1610.07585}}.

\bibitem{Aad:2012awa}
{\bf ATLAS} Collaboration, G.~Aad et~al., {\it {Measurement of $ZZ$ production
  in $pp$ collisions at $\sqrt{s}=7\,$TeV and limits on anomalous $ZZZ$ and
  $ZZ\gamma$ couplings with the ATLAS detector}},  {\em JHEP} {\bf 03} (2013)
  128, [\href{http://arxiv.org/abs/1211.6096}{{\tt arXiv:1211.6096}}].

\bibitem{CMS:2014xja}
{\bf CMS} Collaboration, V.~Khachatryan et~al., {\it {Measurement of the $pp
  \to ZZ$ production cross section and constraints on anomalous triple gauge
  couplings in four-lepton final states at $\sqrt{s}=8\,$TeV}},  {\em Phys.
  Lett.} {\bf B740} (2015) 250--272,
  [\href{http://arxiv.org/abs/1406.0113}{{\tt arXiv:1406.0113}}].

\bibitem{Khachatryan:2015pba}
{\bf CMS} Collaboration, V.~Khachatryan et~al., {\it {Measurements of the $ZZ$
  production cross sections in the $2\ell2\nu$ channel in proton-proton
  collisions at $\sqrt{s} = 7$ and $8\,$TeV and combined constraints on triple
  gauge couplings}},  {\em Eur. Phys. J.} {\bf C75} (2015), no.~10 511,
  [\href{http://arxiv.org/abs/1503.05467}{{\tt arXiv:1503.05467}}].

\bibitem{Aad:2015zqe}
{\bf ATLAS} Collaboration, G.~Aad et~al., {\it {Measurement of the $ZZ$
  Production Cross Section in $pp$ Collisions at $\sqrt{s}$ = 13 TeV with the
  ATLAS Detector}},  {\em Phys. Rev. Lett.} {\bf 116} (2016), no.~10 101801,
  [\href{http://arxiv.org/abs/1512.05314}{{\tt arXiv:1512.05314}}].

\bibitem{Khachatryan:2016txa}
{\bf CMS} Collaboration, V.~Khachatryan et~al., {\it {Measurement of the $ZZ$
  production cross section and $Z \to \ell^+ \ell^- \ell^{\prime +}
  \ell^{\prime -}$ branching fraction in $pp$ collisions at $\sqrt{s} =$ 13
  TeV}},  {\em Submitted to: Phys. Lett. B} (2016)
  [\href{http://arxiv.org/abs/1607.08834}{{\tt arXiv:1607.08834}}].

\bibitem{Ohnemus:1990za}
J.~Ohnemus and J.~F. Owens, {\it {An order $\alpha_{s}$ calculation of hadronic
  $Z Z$ production}},  {\em Phys. Rev.} {\bf D43} (1991) 3626--3639.

\bibitem{Mele:1990bq}
B.~Mele, P.~Nason, and G.~Ridolfi, {\it {QCD radiative corrections to Z boson
  pair production in hadronic collisions}},  {\em Nucl. Phys.} {\bf B357}
  (1991) 409--438.

\bibitem{Ohnemus:1994ff}
J.~Ohnemus, {\it {Hadronic $Z Z$, $W^{-} W^{+}$, and $W^\pm Z$ production with
  QCD corrections and leptonic decays}},  {\em Phys. Rev.} {\bf D50} (1994)
  1931--1945, [\href{http://arxiv.org/abs/hep-ph/9403331}{{\tt
  hep-ph/9403331}}].

\bibitem{Dixon:1998py}
L.~J. Dixon, Z.~Kunszt, and A.~Signer, {\it {Helicity amplitudes for
  $\mathcal{O}(\alpha_s)$ production of $W^{+} W^{-}$, $W^\pm Z$, $Z Z$, $W^\pm
  \gamma$, or $Z \gamma$ pairs at hadron colliders}},  {\em Nucl. Phys.} {\bf
  B531} (1998) 3--23, [\href{http://arxiv.org/abs/hep-ph/9803250}{{\tt
  hep-ph/9803250}}].

\bibitem{Campbell:1999ah}
J.~M. Campbell and R.~K. Ellis, {\it {An update on vector boson pair production
  at hadron colliders}},  {\em Phys. Rev.} {\bf D60} (1999) 113006,
  [\href{http://arxiv.org/abs/hep-ph/9905386}{{\tt hep-ph/9905386}}].

\bibitem{Dixon:1999di}
L.~J. Dixon, Z.~Kunszt, and A.~Signer, {\it {Vector boson pair production in
  hadronic collisions at order $\alpha_s$ : Lepton correlations and anomalous
  couplings}},  {\em Phys. Rev.} {\bf D60} (1999) 114037,
  [\href{http://arxiv.org/abs/hep-ph/9907305}{{\tt hep-ph/9907305}}].

\bibitem{Dicus:1987dj}
D.~A. Dicus, C.~Kao, and W.~W. Repko, {\it {Gluon production of gauge bosons}},
   {\em Phys. Rev.} {\bf D36} (1987) 1570.

\bibitem{Glover:1988rg}
E.~W.~N. Glover and J.~J. van~der Bij, {\it {Z boson pair production via gluon
  fusion}},  {\em Nucl. Phys.} {\bf B321} (1989) 561.

\bibitem{Matsuura:1991pj}
T.~Matsuura and J.~J. van~der Bij, {\it {Characteristics of leptonic signals
  for Z boson pairs at hadron colliders}},  {\em Z. Phys.} {\bf C51} (1991)
  259--266.

\bibitem{Zecher:1994kb}
C.~Zecher, T.~Matsuura, and J.~J. van~der Bij, {\it {Leptonic signals from
  off-shell Z boson pairs at hadron colliders}},  {\em Z. Phys.} {\bf C64}
  (1994) 219--226, [\href{http://arxiv.org/abs/hep-ph/9404295}{{\tt
  hep-ph/9404295}}].

\bibitem{Binoth:2008pr}
T.~Binoth, N.~Kauer, and P.~Mertsch, {\it {Gluon-induced QCD corrections to
  $pp\to ZZ\to l \bar{l} l^\prime \bar{l}^\prime$}},  in {\em {Proceedings,
  16th International Workshop on Deep Inelastic Scattering and Related Subjects
  (DIS 2008)}}, p.~142, 2008.
\newblock \href{http://arxiv.org/abs/0807.0024}{{\tt arXiv:0807.0024}}.

\bibitem{Nason:2006hfa}
P.~Nason and G.~Ridolfi, {\it {A positive-weight next-to-leading-order Monte
  Carlo for Z pair hadroproduction}},  {\em JHEP} {\bf 08} (2006) 077,
  [\href{http://arxiv.org/abs/hep-ph/0606275}{{\tt hep-ph/0606275}}].

\bibitem{Hamilton:2010mb}
K.~Hamilton, {\it {A positive-weight next-to-leading order simulation of weak
  boson pair production}},  {\em JHEP} {\bf 01} (2011) 009,
  [\href{http://arxiv.org/abs/1009.5391}{{\tt arXiv:1009.5391}}].

\bibitem{Hoche:2010pf}
S.~H{\"o}che, F.~Krauss, M.~Sch{\"o}nherr, and F.~Siegert, {\it {Automating the
  POWHEG method in Sherpa}},  {\em JHEP} {\bf 04} (2011) 024,
  [\href{http://arxiv.org/abs/1008.5399}{{\tt arXiv:1008.5399}}].

\bibitem{Melia:2011tj}
T.~Melia, P.~Nason, R.~R{\"o}ntsch, and G.~Zanderighi, {\it {$W^+W^-$, $WZ$ and
  $ZZ$ production in the POWHEG BOX}},  {\em JHEP} {\bf 11} (2011) 078,
  [\href{http://arxiv.org/abs/1107.5051}{{\tt arXiv:1107.5051}}].

\bibitem{Frederix:2011ss}
R.~Frederix, et~al., {\it {Four-lepton production at hadron colliders: aMC@NLO
  predictions with theoretical uncertainties}},  {\em JHEP} {\bf 02} (2012)
  099, [\href{http://arxiv.org/abs/1110.4738}{{\tt arXiv:1110.4738}}].

\bibitem{Cascioli:2013gfa}
F.~Cascioli, et~al., {\it {Precise Higgs-background predictions: merging NLO
  QCD and squared quark-loop corrections to four-lepton + 0,1 jet production}},
   {\em JHEP} {\bf 01} (2014) 046, [\href{http://arxiv.org/abs/1309.0500}{{\tt
  arXiv:1309.0500}}].

\bibitem{Cascioli:2014yka}
F.~Cascioli, et~al., {\it {ZZ production at hadron colliders in NNLO QCD}},
  {\em Phys. Lett.} {\bf B735} (2014) 311--313,
  [\href{http://arxiv.org/abs/1405.2219}{{\tt arXiv:1405.2219}}].

\bibitem{Grazzini:2015hta}
M.~Grazzini, S.~Kallweit, and D.~Rathlev, {\it {ZZ production at the LHC:
  fiducial cross sections and distributions in NNLO QCD}},  {\em Phys. Lett.}
  {\bf B750} (2015) 407--410, [\href{http://arxiv.org/abs/1507.06257}{{\tt
  arXiv:1507.06257}}].

\bibitem{Caola:2015psa}
F.~Caola, K.~Melnikov, R.~Röntsch, and L.~Tancredi, {\it {QCD corrections to
  ZZ production in gluon fusion at the LHC}},  {\em Phys. Rev.} {\bf D92}
  (2015), no.~9 094028, [\href{http://arxiv.org/abs/1509.06734}{{\tt
  arXiv:1509.06734}}].

\bibitem{Caola:2016trd}
F.~Caola, M.~Dowling, K.~Melnikov, R.~Röntsch, and L.~Tancredi, {\it {QCD
  corrections to vector boson pair production in gluon fusion including
  interference effects with off-shell Higgs at the LHC}},  {\em JHEP} {\bf 07}
  (2016) 087, [\href{http://arxiv.org/abs/1605.04610}{{\tt arXiv:1605.04610}}].

\bibitem{Alioli:2016xab}
S.~Alioli, F.~Caola, G.~Luisoni, and R.~R{\"o}ntsch, {\it {ZZ production in
  gluon fusion at NLO matched to parton-shower}},
  \href{http://arxiv.org/abs/1609.09719}{{\tt arXiv:1609.09719}}.

\bibitem{Beenakker:1993tt}
W.~Beenakker, A.~Denner, S.~Dittmaier, R.~Mertig, and T.~Sack, {\it
  {High-energy approximation for on-shell W pair production}},  {\em
  Nucl.Phys.} {\bf B410} (1993) 245--279.

\bibitem{Beccaria:1998qe}
M.~Beccaria, G.~Montagna, F.~Piccinini, F.~Renard, and C.~Verzegnassi, {\it
  {Rising bosonic electroweak virtual effects at high-energy $e^+ e^-$
  colliders}},  {\em Phys.Rev.} {\bf D58} (1998) 093014,
  [\href{http://arxiv.org/abs/hep-ph/9805250}{{\tt hep-ph/9805250}}].

\bibitem{Ciafaloni:1998xg}
P.~Ciafaloni and D.~Comelli, {\it {Sudakov enhancement of electroweak
  corrections}},  {\em Phys.Lett.} {\bf B446} (1999) 278--284,
  [\href{http://arxiv.org/abs/hep-ph/9809321}{{\tt hep-ph/9809321}}].

\bibitem{Kuhn:1999de}
J.~H. K{\"u}hn and A.~Penin, {\it {Sudakov logarithms in electroweak
  processes}},  \href{http://arxiv.org/abs/hep-ph/9906545}{{\tt
  hep-ph/9906545}}.

\bibitem{Fadin:1999bq}
V.~S. Fadin, L.~N. Lipatov, A.~D. Martin, and M.~Melles, {\it {Resummation of
  double logarithms in electroweak high-energy processes}},  {\em Phys. Rev.}
  {\bf D61} (2000) 094002, [\href{http://arxiv.org/abs/hep-ph/9910338}{{\tt
  hep-ph/9910338}}].

\bibitem{Denner:2000jv}
A.~Denner and S.~Pozzorini, {\it {One loop leading logarithms in electroweak
  radiative corrections. 1. Results}},  {\em Eur.Phys.J.} {\bf C18} (2001)
  461--480, [\href{http://arxiv.org/abs/hep-ph/0010201}{{\tt hep-ph/0010201}}].

\bibitem{Accomando:2004de}
E.~Accomando, A.~Denner, and A.~Kaiser, {\it {Logarithmic electroweak
  corrections to gauge-boson pair production at the LHC}},  {\em Nucl. Phys.}
  {\bf B706} (2005) 325--371, [\href{http://arxiv.org/abs/hep-ph/0409247}{{\tt
  hep-ph/0409247}}].

\bibitem{Bierweiler:2013dja}
A.~Bierweiler, T.~Kasprzik, and J.~H. K{\"u}hn, {\it {Vector-boson pair
  production at the LHC to $\mathcal{O}(\alpha^3)$ accuracy}},  {\em JHEP} {\bf
  12} (2013) 071, [\href{http://arxiv.org/abs/1305.5402}{{\tt
  arXiv:1305.5402}}].

\bibitem{Baglio:2013toa}
J.~Baglio, L.~D. Ninh, and M.~M. Weber, {\it {Massive gauge boson pair
  production at the LHC: a next-to-leading order story}},  {\em Phys. Rev.}
  {\bf D88} (2013) 113005, [\href{http://arxiv.org/abs/1307.4331}{{\tt
  arXiv:1307.4331}}].

\bibitem{Billoni:2013aba}
M.~Billoni, S.~Dittmaier, B.~J{\"a}ger, and C.~Speckner, {\it {Next-to-leading
  order electroweak corrections to $pp\to W^+W^- \to 4$ leptons at the LHC in
  double-pole approximation}},  {\em JHEP} {\bf 12} (2013) 043,
  [\href{http://arxiv.org/abs/1310.1564}{{\tt arXiv:1310.1564}}].

\bibitem{Gieseke:2014gka}
S.~Gieseke, T.~Kasprzik, and J.~H. Kühn, {\it {Vector-boson pair production
  and electroweak corrections in HERWIG++}},  {\em Eur. Phys. J.} {\bf C74}
  (2014), no.~8 2988, [\href{http://arxiv.org/abs/1401.3964}{{\tt
  arXiv:1401.3964}}].

\bibitem{Biedermann:2016guo}
B.~Biedermann, et~al., {\it {Next-to-leading-order electroweak corrections to
  $pp \to W^+W^-\to4$~leptons at the LHC}},  {\em JHEP} {\bf 06} (2016) 065,
  [\href{http://arxiv.org/abs/1605.03419}{{\tt arXiv:1605.03419}}].

\bibitem{Biedermann:2016yvs}
B.~Biedermann, A.~Denner, S.~Dittmaier, L.~Hofer, and B.~Jäger, {\it
  {Electroweak corrections to $pp \to \mu^+\mu^-e^+e^- + X$ at the LHC: a Higgs
  background study}},  {\em Phys. Rev. Lett.} {\bf 116} (2016), no.~16 161803,
  [\href{http://arxiv.org/abs/1601.07787}{{\tt arXiv:1601.07787}}].

\bibitem{Denner:1999gp}
A.~Denner, S.~Dittmaier, M.~Roth, and D.~Wackeroth, {\it {Predictions for all
  processes $e^+ e^- \to 4\,$fermions${}+\gamma$}},  {\em Nucl.Phys.} {\bf
  B560} (1999) 33--65, [\href{http://arxiv.org/abs/hep-ph/9904472}{{\tt
  hep-ph/9904472}}].

\bibitem{Denner:2005fg}
A.~Denner, S.~Dittmaier, M.~Roth, and L.~Wieders, {\it {Electroweak corrections
  to charged-current $e^+ e^- \to 4$ fermion processes: Technical details and
  further results}},  {\em Nucl.Phys.} {\bf B724} (2005) 247--294,
  [\href{http://arxiv.org/abs/hep-ph/0505042}{{\tt hep-ph/0505042}}].

\bibitem{Denner:2006ic}
A.~Denner and S.~Dittmaier, {\it {The complex-mass scheme for perturbative
  calculations with unstable particles}},  {\em Nucl.Phys.Proc.Suppl.} {\bf
  160} (2006) 22--26, [\href{http://arxiv.org/abs/hep-ph/0605312}{{\tt
  hep-ph/0605312}}].

\bibitem{Piguet:1974tp}
O.~Piguet, {\it {Construction of a strictly renormalizable effective Lagrangian
  for the massive Abelian Higgs model}},  {\em Commun. Math. Phys.} {\bf 37}
  (1974) 19.

\bibitem{Collins:1984xc}
J.~C. Collins, {\em {Renormalization}}, vol.~26 of {\em Cambridge Monographs on
  Mathematical Physics}.
\newblock Cambridge University Press, Cambridge, 1986.

\bibitem{Catani:1996vz}
S.~Catani and M.~Seymour, {\it {A general algorithm for calculating jet
  cross-sections in NLO QCD}},  {\em Nucl.Phys.} {\bf B485} (1997) 291--419,
  [\href{http://arxiv.org/abs/hep-ph/9605323}{{\tt hep-ph/9605323}}].

\bibitem{Dittmaier:1999mb}
S.~Dittmaier, {\it {A general approach to photon radiation off fermions}},
  {\em Nucl.Phys.} {\bf B565} (2000) 69--122,
  [\href{http://arxiv.org/abs/hep-ph/9904440}{{\tt hep-ph/9904440}}].

\bibitem{Aad:2011gj}
{\bf ATLAS Collaboration} Collaboration, G.~Aad et~al., {\it {Measurement of
  the transverse momentum distribution of $\PZ/\gamma^*$ bosons in
  proton-proton collisions at $\sqrt{s}=7$ TeV with the ATLAS detector}},  {\em
  Phys.Lett.} {\bf B705} (2011) 415--434,
  [\href{http://arxiv.org/abs/1107.2381}{{\tt arXiv:1107.2381}}].

\bibitem{Dittmaier:2008md}
S.~Dittmaier, A.~Kabelschacht, and T.~Kasprzik, {\it {Polarized QED splittings
  of massive fermions and dipole subtraction for non-collinear-safe
  observables}},  {\em Nucl.Phys.} {\bf B800} (2008) 146--189,
  [\href{http://arxiv.org/abs/0802.1405}{{\tt arXiv:0802.1405}}].

\bibitem{Dittmaier:2009cr}
S.~Dittmaier and M.~Huber, {\it {Radiative corrections to the neutral-current
  Drell-Yan process in the Standard Model and its minimal supersymmetric
  extension}},  {\em JHEP} {\bf 01} (2010) 060,
  [\href{http://arxiv.org/abs/0911.2329}{{\tt arXiv:0911.2329}}].

\bibitem{Actis:2016mpe}
S.~Actis, et~al., {\it {RECOLA: REcursive Computation of One-Loop Amplitudes}},
   \href{http://arxiv.org/abs/1605.01090}{{\tt arXiv:1605.01090}}.

\bibitem{Catani:2002hc}
S.~Catani, S.~Dittmaier, M.~H. Seymour, and Z.~Trocsanyi, {\it {The dipole
  formalism for next-to-leading order QCD calculations with massive partons}},
  {\em Nucl.Phys.} {\bf B627} (2002) 189--265,
  [\href{http://arxiv.org/abs/hep-ph/0201036}{{\tt hep-ph/0201036}}].

\bibitem{Accomando:2005ra}
E.~Accomando, A.~Denner, and C.~Meier, {\it {Electroweak corrections to $W
  \gamma$ and $Z \gamma$ production at the LHC}},  {\em Eur. Phys. J.} {\bf
  C47} (2006) 125--146, [\href{http://arxiv.org/abs/hep-ph/0509234}{{\tt
  hep-ph/0509234}}].

\bibitem{Kublbeck:1990xc}
J.~{K\"ublbeck}, M.~{B\"ohm}, and A.~Denner, {\it {FeynArts: Computer algebraic
  generation of Feynman graphs and amplitudes}},  {\em Comput. Phys. Commun.}
  {\bf 60} (1990) 165--180.

\bibitem{Hahn:2000kx}
T.~Hahn, {\it {Generating Feynman diagrams and amplitudes with FeynArts 3}},
  {\em Comput. Phys. Commun.} {\bf 140} (2001) 418--431,
  [\href{http://arxiv.org/abs/hep-ph/0012260}{{\tt hep-ph/0012260}}].

\bibitem{Hahn:1998yk}
T.~Hahn and M.~Perez-Victoria, {\it {Automatized one loop calculations in
  four-dimensions and D-dimensions}},  {\em Comput. Phys. Commun.} {\bf 118}
  (1999) 153--165, [\href{http://arxiv.org/abs/hep-ph/9807565}{{\tt
  hep-ph/9807565}}].

\bibitem{Denner:2005es}
A.~Denner, S.~Dittmaier, M.~Roth, and L.~Wieders, {\it {Complete electroweak
  $\mathcal{O}(\alpha)$ corrections to charged-current $e^+e^-\to4\,$ fermion
  processes}},  {\em Phys.Lett.} {\bf B612} (2005) 223--232,
  [\href{http://arxiv.org/abs/hep-ph/0502063}{{\tt hep-ph/0502063}}].

\bibitem{Denner:2016kdg}
A.~Denner, S.~Dittmaier, and L.~Hofer, {\it {Collier: a fortran-based Complex
  One-Loop LIbrary in Extended Regularizations}},
  \href{http://arxiv.org/abs/1604.06792}{{\tt arXiv:1604.06792}}.

\bibitem{Denner:2002ii}
A.~Denner and S.~Dittmaier, {\it {Reduction of one loop tensor five-point
  integrals}},  {\em Nucl.Phys.} {\bf B658} (2003) 175--202,
  [\href{http://arxiv.org/abs/hep-ph/0212259}{{\tt hep-ph/0212259}}].

\bibitem{Denner:2005nn}
A.~Denner and S.~Dittmaier, {\it {Reduction schemes for one-loop tensor
  integrals}},  {\em Nucl.Phys.} {\bf B734} (2006) 62--115,
  [\href{http://arxiv.org/abs/hep-ph/0509141}{{\tt hep-ph/0509141}}].

\bibitem{Beenakker:1990jr}
W.~Beenakker and A.~Denner, {\it {Infrared divergent scalar box integrals with
  applications in the Electroweak Standard Model}},  {\em Nucl.Phys.} {\bf
  B338} (1990) 349--370.

\bibitem{Denner:1991qq}
A.~Denner, U.~Nierste, and R.~Scharf, {\it {A compact expression for the scalar
  one loop four point function}},  {\em Nucl.Phys.} {\bf B367} (1991) 637--656.

\bibitem{Denner:2010tr}
A.~Denner and S.~Dittmaier, {\it {Scalar one-loop 4-point integrals}},  {\em
  Nucl.Phys.} {\bf B844} (2011) 199--242,
  [\href{http://arxiv.org/abs/1005.2076}{{\tt arXiv:1005.2076}}].

\bibitem{Berends:1994pv}
F.~A. Berends, R.~Pittau, and R.~Kleiss, {\it {All electroweak four fermion
  processes in electron--positron collisions}},  {\em Nucl. Phys.} {\bf B424}
  (1994) 308--342, [\href{http://arxiv.org/abs/hep-ph/9404313}{{\tt
  hep-ph/9404313}}].

\bibitem{Dittmaier:2002ap}
S.~Dittmaier and M.~Roth, {\it {LUSIFER: A LUcid approach to six FERmion
  production}},  {\em Nucl. Phys.} {\bf B642} (2002) 307--343,
  [\href{http://arxiv.org/abs/hep-ph/0206070}{{\tt hep-ph/0206070}}].

\bibitem{Motz:Thesis}
{Tobias Motz}, {\it {Generic Monte Carlo Event Generator for NLO Processes}},
  {Dissertation}, {Universit\"at Z\"urich}, {2011}.

\bibitem{Bardin:1988xt}
D.~{\relax Yu}. Bardin, A.~Leike, T.~Riemann, and M.~Sachwitz, {\it
  {Energy-dependent width effects in $e^+ e^-$ annihilation near the Z-boson
  pole}},  {\em Phys. Lett.} {\bf B206} (1988) 539--542.

\bibitem{Ball:2013hta}
{\bf NNPDF} Collaboration, R.~D. Ball, et~al., {\it {Parton distributions with
  QED corrections}},  {\em Nucl. Phys.} {\bf B877} (2013) 290--320,
  [\href{http://arxiv.org/abs/1308.0598}{{\tt arXiv:1308.0598}}].

\bibitem{Manohar:2016nzj}
A.~Manohar, P.~Nason, G.~P. Salam, and G.~Zanderighi, {\it {How bright is the
  proton? A precise determination of the photon PDF}},
  \href{http://arxiv.org/abs/1607.04266}{{\tt arXiv:1607.04266}}.

\bibitem{Diener:2005me}
K.-P. Diener, S.~Dittmaier, and W.~Hollik, {\it {Electroweak higher-order
  effects and theoretical uncertainties in deep-inelastic neutrino
  scattering}},  {\em Phys.Rev.} {\bf D72} (2005) 093002,
  [\href{http://arxiv.org/abs/hep-ph/0509084}{{\tt hep-ph/0509084}}].

\bibitem{Aad:2014eva}
{\bf ATLAS} Collaboration, G.~Aad et~al., {\it {Measurements of Higgs boson
  production and couplings in the four-lepton channel in pp collisions at
  center-of-mass energies of $7$ and $8\,$TeV with the ATLAS detector}},  {\em
  Phys.Rev.} {\bf D91} (2015), no.~1 012006,
  [\href{http://arxiv.org/abs/1408.5191}{{\tt arXiv:1408.5191}}].

\bibitem{Chatrchyan:2013mxa}
{\bf CMS} Collaboration, S.~Chatrchyan et~al., {\it {Measurement of the
  properties of a Higgs boson in the four-lepton final state}},  {\em Phys.
  Rev.} {\bf D89} (2014), no.~9 092007,
  [\href{http://arxiv.org/abs/1312.5353}{{\tt arXiv:1312.5353}}].

\bibitem{Denner:2015fca}
A.~Denner, S.~Dittmaier, M.~Hecht, and C.~Pasold, {\it {NLO QCD and electroweak
  corrections to $Z+\gamma$ production with leptonic Z-boson decays}},  {\em
  JHEP} {\bf 02} (2016) 057, [\href{http://arxiv.org/abs/1510.08742}{{\tt
  arXiv:1510.08742}}].

\bibitem{Bredenstein:2006rh}
A.~Bredenstein, A.~Denner, S.~Dittmaier, and M.~M. Weber, {\it {Precise
  predictions for the Higgs-boson decay H $\to$ WW/ZZ $\to$ 4 leptons}},  {\em
  Phys. Rev.} {\bf D74} (2006) 013004,
  [\href{http://arxiv.org/abs/hep-ph/0604011}{{\tt hep-ph/0604011}}].

\bibitem{Bredenstein:2007ec}
A.~Bredenstein, A.~Denner, S.~Dittmaier, and M.~M. Weber, {\it {Precision
  calculations for H $\to$ WW/ZZ $\to$ 4fermions with PROPHECY4f}},  in {\em
  {Proceedings, International Linear Collider Workshop (LCWS07 and ILC07):
  Hamburg , Germany, May 30-June 3, 2007, Vol.1-2}}, pp.~150--154, 2007.
\newblock \href{http://arxiv.org/abs/0708.4123}{{\tt arXiv:0708.4123}}.

\end{thebibliography}\endgroup

\end{document}